\documentclass[preprintnumbers,amsmath,amssymb,superscriptaddress,twocolumn,showpacs]{revtex4-2}
\usepackage{graphicx}
\usepackage{dcolumn}
\usepackage{bm}
\usepackage{physics}
\usepackage{ulem}
 \usepackage{xcolor}
\usepackage[caption=false]{subfig}
\usepackage[english]{babel}
\usepackage{xr-hyper}
\usepackage{multirow}
\usepackage{booktabs}
\usepackage{array}
\usepackage{siunitx}
\usepackage[hidelinks]{hyperref}

\usepackage{cancel}
\newcommand{\be}{\begin{equation}}
\newcommand{\ee}{\end{equation}}
\newcommand{\bea}{\begin{eqnarray}}
\newcommand{\eea}{\end{eqnarray}}

\begin{document}

\title{High-Sensitivity Optical Detection of Electron-Nuclear Spin Clusters in Diamond}

\author{L. Chambard}
\affiliation{Laboratoire De Physique de l'\'Ecole Normale Sup\'erieure, \'Ecole Normale Sup\'erieure, PSL Research University, CNRS, Sorbonne Universit\'e, Universit\'e Paris Cit\'e , 24 rue Lhomond, 75231 Paris Cedex 05, France}

\author{A. Durand}
\affiliation{Laboratoire De Physique de l'\'Ecole Normale Sup\'erieure, \'Ecole Normale Sup\'erieure, PSL Research University, CNRS, Sorbonne Universit\'e, Universit\'e Paris Cit\'e , 24 rue Lhomond, 75231 Paris Cedex 05, France}

\author{J. Voisin}
\affiliation{Laboratoire De Physique de l'\'Ecole Normale Sup\'erieure, \'Ecole Normale Sup\'erieure, PSL Research University, CNRS, Sorbonne Universit\'e, Universit\'e Paris Cit\'e , 24 rue Lhomond, 75231 Paris Cedex 05, France}

\author{M. Perdriat}
\affiliation{Laboratoire De Physique de l'\'Ecole Normale Sup\'erieure, \'Ecole Normale Sup\'erieure, PSL Research University, CNRS, Sorbonne Universit\'e, Universit\'e Paris Cit\'e , 24 rue Lhomond, 75231 Paris Cedex 05, France}

\author{V. Jacques}
\affiliation{Laboratoire Charles Coulomb, Université de Montpellier, CNRS, Montpellier, France}

\author{G. H\'etet} 
\affiliation{Laboratoire De Physique de l'\'Ecole Normale Sup\'erieure, \'Ecole Normale Sup\'erieure, PSL Research University, CNRS, Sorbonne Universit\'e, Universit\'e Paris Cit\'e , 24 rue Lhomond, 75231 Paris Cedex 05, France}

\begin{abstract}
We perform sensitive nuclear magnetic resonance (NMR) with spin ensembles which are polarized by nitrogen vacancy centers (NV centers) in diamond at room-temperature. Using {stabilized lasers, a balanced detection}  and a highly uniform magnetic field, we resolve sharp NMR features arising from multiple spin clusters. In particular, we investigate the coupling between nuclear spins and NV centers in the neutral and negatively charged states. Further, we perform high precision NMR and coherent control of families of $^{13}$C nuclear spin ensembles in the $\ket{m_s=0}$ level of the NV ground state.
Applying an off-axis magnetic field reveals the various sites associated with the otherwise degenerate couplings of the $^{13}$C sites around the NV electronic spin providing access to all the hyperfine tensor components. Last, we observe spectroscopic signatures of pairs of nuclear spins coupled to the same NV center.  These results are relevant for ensemble measurements of dynamical polarization that currently rely on expensive nuclear magnetic resonance systems as well as for recently proposed nuclear spin gyroscopes.
\end{abstract}

\maketitle

Harnessing ensembles of nuclear spins lies at the heart of a vibrant interdisciplinary effort, driven by applications in quantum information processing, secure communication, and nanoscale sensing \cite{Appel2024, Marcks, Kenny}. Among the most promising platforms in this field of research is the negatively charged nitrogen-vacancy (NV$^-$) center in diamond, whose electronic spin state allows for efficient readout and polarization of proximal nuclear spins via optical means at room temperature. Notable breakthroughs using this hybrid electronic-nuclear platform include the realization of large-scale nuclear spin registers \cite{Bradley}, high-precision gyroscopy \cite{ledbetter2012gyroscopes, Ajoy2012, Jarmola}, nanoscale Nuclear Magnetic Resonance (NMR) spectroscopy in microfluidic environments \cite{Smits, Tetienne} or Nanoscale Magnetic Resonance imaging  \cite{Budakian_2024}. Despite the unique ability of NV centers to detect nuclear spins optically, ensemble-based detection of bulk nuclear polarization—especially for $^{13}$C spins—still often relies on conventional NMR setups, which require strong (Tesla-scale) magnetic fields and cryogenic temperatures \cite{Ajoy2018, Scheuer, Henshaw, Blinder}.

A promising alternative leverages optical detection of nuclear magnetic resonance (ODNMR), particularly {\it via} the nitrogen nuclear spin of the NV center itself. In this method, nuclear polarization is enhanced at the excited-state level anti-crossing (ESLAC), where the NV photoluminescence becomes sensitive to the nuclear spin state \cite{smeltzer_pra, Steiner, Jarmola_1}. While $^{14}$N nuclei are commonly used in this context, the detection of other spin ensembles, such as $^{13}$C, remains more challenging due, notably, to reduced polarization efficiency caused by anisotropic hyperfine fields in the NV optically excited state and the inhomogenous hyperfine coupling of the distant nuclear spins to the NV electronic spin. To date, most high-resolution studies of $^{13}$C nuclear spins have relied on the probabilistic study of several single NV centers \cite{Childress2006, Dreau2012}, which restricts throughput and scalability. Bulk readout of ensembles of $^{13}$C nuclear spins generally still require conventional NMR techniques.
In the present work, we perform ODNMR {with NV ensembles} using amplitude and beam direction-stabilized, high-power green laser excitation in conjunction with balanced photoluminescence detection and highly homogeneous magnetic field conditions. By operating near the ESLAC, we achieve sensitive and high resolution detection of weak nuclear spin signals. 
Our results show that even in a disordered system where all coupled $^{13}$C-NV pairs are present, discrete families can be resolved. In particular, we detect distinct $^{13}$C families (labeled A, B, C and D) in the NV electronic ground state manifold as well as site degeneracy lifting using a novel two-tones correlation spectroscopy method. 
We also report the optical detection of nuclear spins of the $^{14}$N atom in the neutral NV$^0$ charge state.
Last, we study pairs of coupled carbon 13 spins that are indirectly connected by the NV center electronic spin.
By driving to triplet states of the coupled system we identify conspicuous features of these pairs of $^{13}$C in good agreement with a numerical model, paving the way for a rich low-field $^{13}$C NMR research landscape.

The measurements in this study are realized with a homemade confocal microscope (see Supplementary Materials \cite{supp}, Section I). Lasers delivering 200 mW in the green (either from Ventus-GEM or a DPSS Cobolt Samba) are employed to excite the NV centers. 
Of central importance to this study is the noise reduction on the detected NV photoluminescence, 
stemming mostly from the laser noise.  
Both lasers were found to be prone to beam pointing and amplitude noise translating directly to PL noise if no correction is applied. 
We reduced the low frequency laser noise thanks to a balanced detection as well as an active correction of laser beam pointing and amplitude noise (see SM, Section I). Such stabilization is crucial for long wide-band NMR scans. 

To minimize magnetic noise from paramagnetic impurities, we use a Chemical-Vapor-Deposition (CVD) grown diamond from the company element 6 (BNV14 diamond) containing 4.5 ppm of NV centers and a natural abundance of $^{13}$C ({\it i.e.} 1.1\%). 
With a laser spot size of about 5 $\mu$m$^2$  at the diamond surface, we detect up to 40 $\mu$W of stable NV photoluminescence. 
Under these power conditions, the PL signal noise—referenced against a green laser—remains at $\sim 10^{-4}\,\mathrm{V}$ for over an hour, enabling efficient NV signal averaging.

\begin{figure}
\includegraphics[width=8cm]{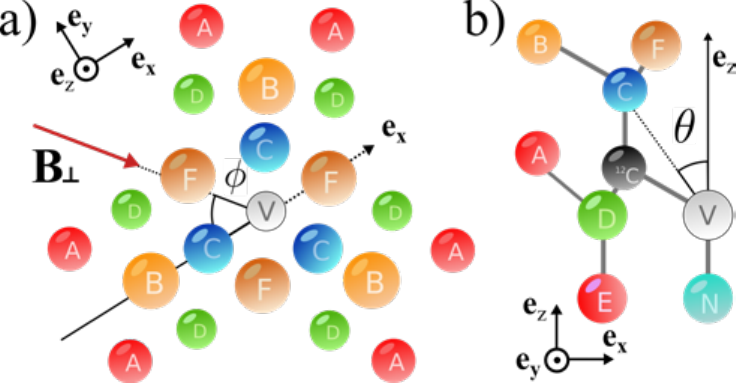}
\caption{ {a) Schematics showing the repartition of $^{13}$C atoms around the NV centers projected on a plane perpendicular to the NV axis. Families are defined according to their symmetries with respect to the NV axis, following the notation introduced in \cite{Smeltzer_njp}. $\phi$ defines the angle between ${\bm B}_\perp$ and the axis $^{13}$C-NV.  b) Schematics showing the $^{13}$C families projected on a plane containing the NV axis. $\theta$ defines the angle between the perpendicular plane passing though the vacancy and $^{13}$C families location.}}
\label{C13}
\end{figure}
The ground electronic state of the negatively charged NV center is a triplet state \cite{DOHERTY20131}.  In the absence of an external magnetic field, the $\vert m_s=\pm 1 \rangle\equiv \ket{\pm 1_e}$ states in the triplet manifold are at a frequency $D \approx 2.87$ GHz above the $\vert m_s=0 \rangle\equiv\ket{0_e}$ state at room temperature. The electronic spin of the NV centers can be optically polarized to the $\ket{0_e}$ state using green laser light. Additionally, the Stokes-shifted photoluminescence (PL) is stronger in the $\ket{0_e}$ state compared to the $\ket{\pm 1_e}$ states. Consequently, applying a microwave signal at the resonant frequency induces a drop in PL, which forms the basis of Optically Detected Magnetic Resonance (ODMR) spectroscopy.
 
\begin{figure}[ht!]
\includegraphics[width=8.5cm]{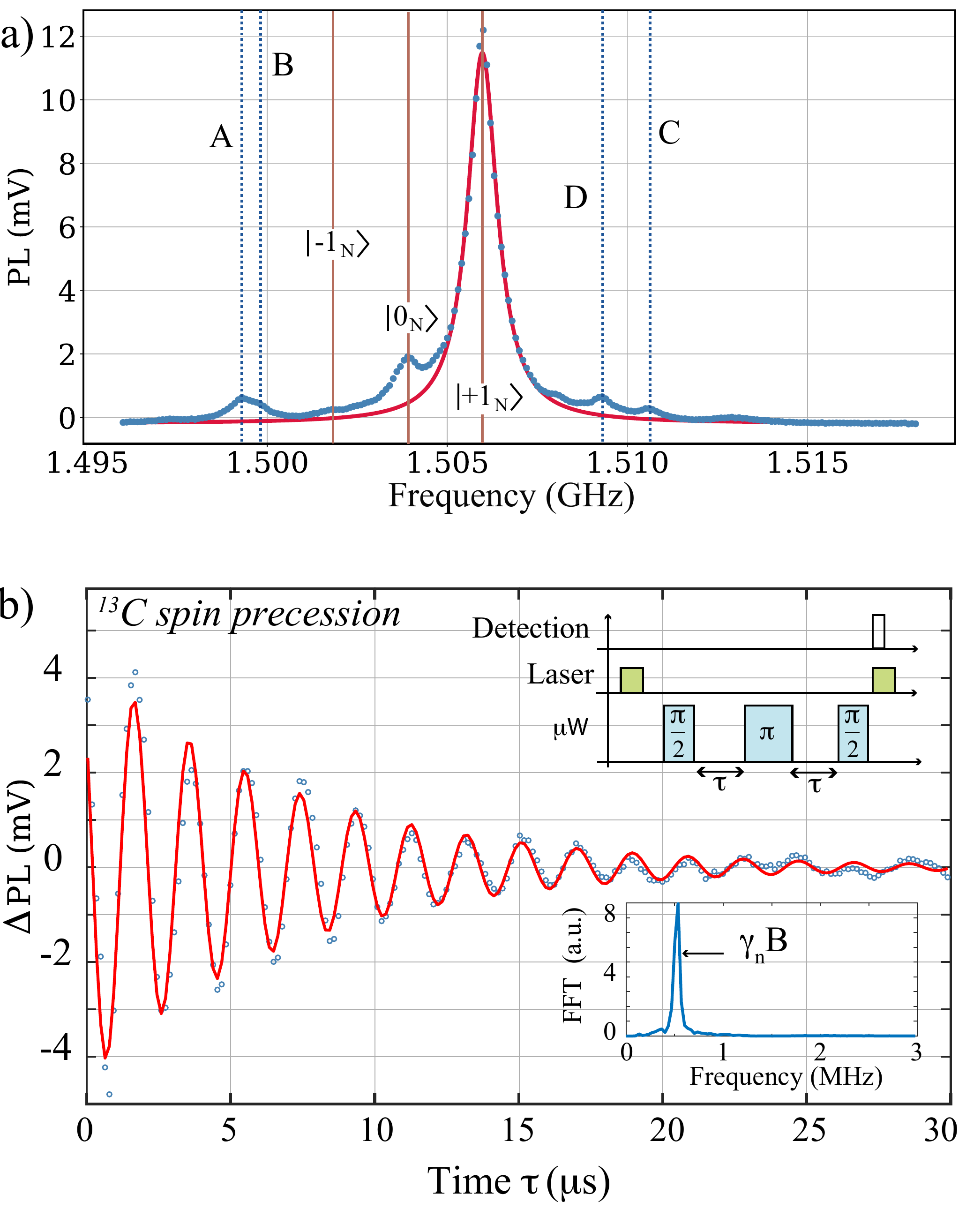}
\caption{a) Optically detected magnetic resonance (ODMR) spectrum
around the $\ket{0_e}\rightarrow\ket{-1_e}$ transition at B= 487~G. Blue points are experimental data, red line is a Lorentzian fit to the resonance associated with the $\ket{0_e, 1_N}\rightarrow \ket{-1_e, 1_N}$ transition. Electronic spin transitions associated with the NV electron coupled to the A, B, C and D $^{13}$C nuclear spin families are indicated by vertical lines. b)  {Electron spin echo envelope modulation signal (ESEEM)} subtracted from the slowly varying envelope (see main text). Insets: Spin echo sequence and Fourier transform of the spin echo curve.}
\label{ODMR}
\end{figure}

To polarize the nuclear spins, a highly homogeneous magnetic field generated by a pair of 2 inches permanent magnets and mounted on a rotation stage is employed (see SM, section II).
We will mostly be studying the $\ket{I=1/2, m_I=\pm 1/2}\equiv \ket{\pm 1/2}$ spin of the $^{13}$C atoms.
Neglecting the coupling between the electronic spin and the nuclear spin of the $^{14}$N, the Hamiltonian that describes a single coupled $^{13}$C-NV pair is given by: \begin{equation}\frac{H}{\hbar} = DS_z^2 + \gamma_e \mathbf{S}\cdot\mathbf{B} + \gamma_n \mathbf{I}\cdot\mathbf{B} + \mathbf{S} \cdot \underline{\underline{\mathbf{A}}} \cdot \mathbf{I},
\end{equation}
where $D= 2.87$~GHz, $\gamma_e=$2.8025~MHz/G and $\gamma_n=-1.07$~kHz/G. $\underline{\underline{\mathbf{A}}}$ is the hyperfine tensor, whose components characterize the anisotropic interaction between the electronic and nuclear spins. In the NV center basis, it is possible to rewrite this Hamiltonian as

\begin{eqnarray}\label{Ham2}
    \frac{H}{\hbar}&=& DS_z^2 + \gamma_e \mathbf{S}\cdot\mathbf{B} + {}\gamma_n \mathbf{I}\cdot\mathbf{B} + S_z I_z A_{\parallel} \nonumber \\
    & +& \frac{A_{\perp}}{2} (S^+ I^- + S^- I^+) \nonumber \\
    &+ &\frac{A_{\text{ani}}}{2} ((S^+ I_z + S_z I^+) e^{-i\phi}
     +  (S^- I_z + S_z I^-) e^{+i\phi}),\nonumber \\
     &&
\end{eqnarray}
where double quantum terms are neglected (See SM, section VI). The $\{\bm e_x,\bm e_y, \bm e_z\}$ coordinate system is shown in Fig.~\ref{C13}-a) and b).
$\phi$ is defined as the angle that points to a $^{13}$C site as depicted in Fig.~\ref{C13}-b). $A_{\rm ani}$, $A_\perp$ are linear combinations of the components of the $\underline{\underline{\mathbf{A}}}$ tensor (see Supplementary materials \cite{supp}, section VI). $A_\perp$ enables a flip-flop process, while $A_{\rm ani}$ couples the electronic and nuclear spins by inducing spin flips of one system conditional on the $z$-projection of the other.

 \begin{figure*}
\includegraphics[width=14cm]{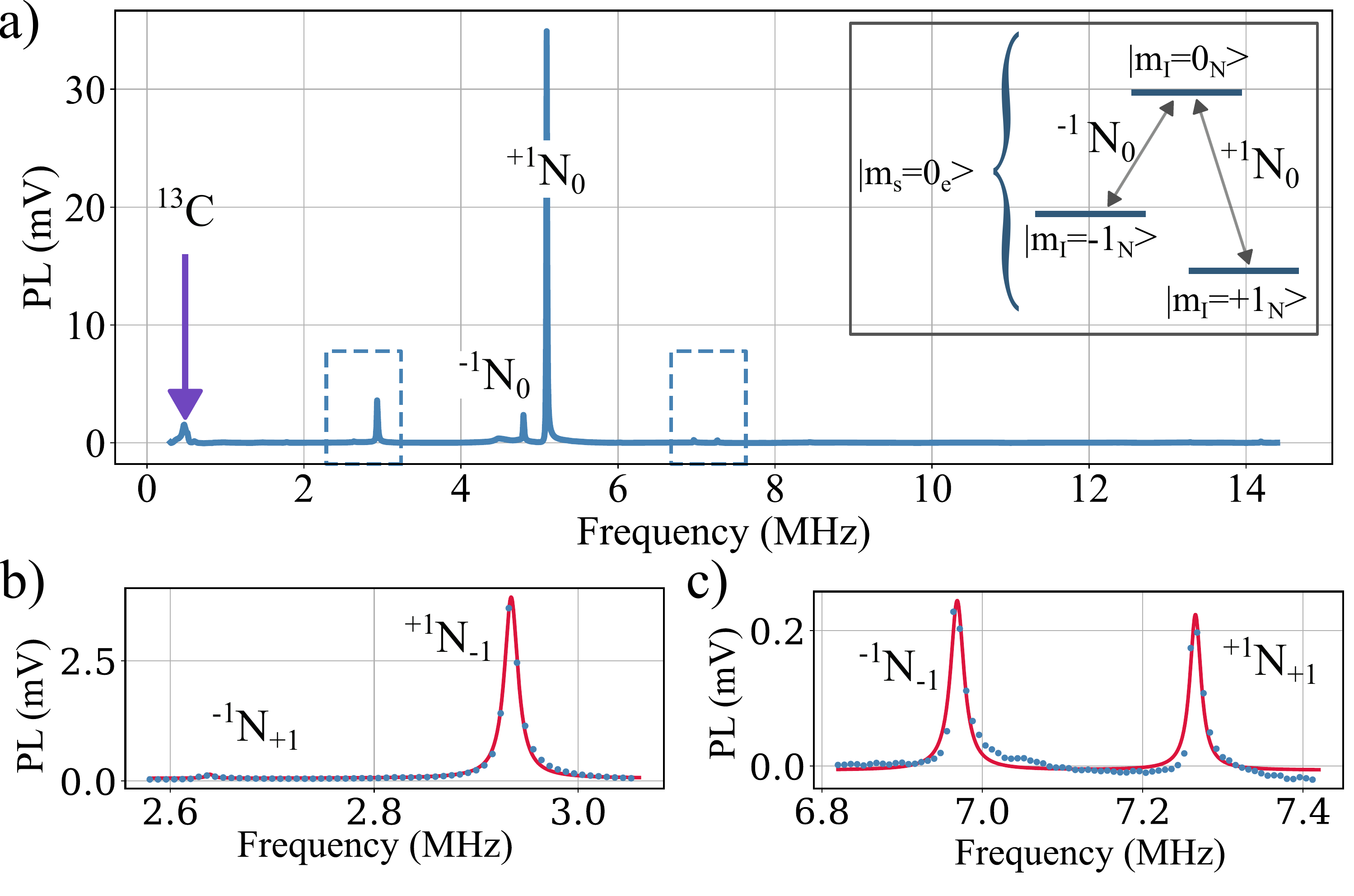}
\caption{a) Optically Detected Nuclear Magnetic Resonance (ODNMR) spectrum at B= 486.8 G. Evolution of the photoluminescence (PL) as a function of the frequency of radio-frequency field, from 400~kHz to 14.3~MHz. The upper right box show the $I_N=1$ level scheme of the   $^{14}$N nuclear spin in the $\ket{0_e}$ state. The $^{13}$C down-arrow shows the expected peak at the $^{13}$C spin Larmor frequency.   b),c) Enlarged spectra corresponding to the two dashed boxes in a), showing resonances associated with the $^{14}$N nuclear spins coupled to NV$^-$ centers in the $\ket{\pm 1_e}$ manifolds.}
\label{ODNMR}
\end{figure*}

At room temperature, the spin Hamiltonian has the same form in both the NV ground and excited states; only the parameters $D$ and $\underline{\underline{\mathbf{A}}}$ differ \cite{DOHERTY20131}.
 In the optically excited state $D^{\rm ex}\approx 1.4$~GHz.
A level crossing between the $\ket{0_e^{\text{ex}}}$ and $\ket{-1^{\text{ex}}}$  takes place at a magnetic field of around 500 G.
At such Excited State Level Anti-crossing (ESLAC), some families of $^{13}$C nuclear spins are polarized because of a joint flipping of the electron and nuclear spins $\ket{0_e^{\text{ex}},-1/2} \rightarrow \ket{-1_e^{\text{ex}},+1/2}$ induced by the perpendicular component $A_\perp^{\text{ex}}$ of the hyperfine tensor. 
After a few optical cycles, at relative angles between NV and magnetic axes below 1 degree, the nuclear spins of atomic species coupled to the NV center with $A_\perp$ larger than a few MHz can then be polarized in one of the $\ket{\pm 1/2}$ states through their hyperfine coupling to the NV electronic spins, at room-temperature \cite{jacques2009}. $^{13}$C families A, B, C, D, up to O have been observed using Optically Detected Magnetic Resonance (ODMR) and partially polarized  using single NV centers in CVD grown samples \cite{Dreau2012,Smeltzer_njp}. 
Fig. \ref{C13}-a) depicts a projected image of the  position of 27 $^{13}$C atoms around the NV centers on a plane perpendicular to the NV axis. 
Fig. \ref{C13}-b) is a schematics showing the $^{13}$C families projected on a plane containing the NV axis. $\theta$ defines the angle between the perpendicular plane passing though the vacancy and $^{13}$C families location.
$\phi$ is the angle between $\bm B_\perp$ and the $^{13}$C-V axis (see Section VI of the supplementary materials \cite{supp}).

The previous discussion focused on $^{13}$C nuclear spins, but ESLAC polarization also applies to  other nuclear species, such as the nuclear spin $I=1$ of the $^{14}$N, provided that the transverse component of the hyperfine tensor in the NV excited state exceeds a few MHz. 

\section{ODMR and Spin Echo at the ESLAC}

Fig.~\ref{ODMR}-a) shows an ODMR scan centered on the $\ket{0_e} \rightarrow \ket{-1_e}$ transition at a homogeneous magnetic field $B=487$~G {aligned to one of the four NV center families} \cite{Fischer}.
{
This value was chosen to maximize nuclear spin coupling to the NV center at the ESLAC \cite{jacques2009}, while ensuring no cross-relaxation with $P_1$ centers \cite{Jarmola_1}.
Magnetic field alignment is achieved using a pair of 2 inches permanent magnets to minimize field gradients. The sample is mounted at the center of a goniometric system, where the field gradient is minimal—on the order of a few $0.06$ T/m along $z$ (see supplemental materials \cite{supp}, section I and II)}.

The main peak at $1506$ MHz corresponds to the electronic spin transition in the polarized $\ket{m_I=+1}\equiv \ket{+1_N}$ nuclear spin state of the $^{14}$N at the 
frequency $\nu\approx D-\gamma_e B + |A_{\parallel}|$. Here, $A_{\parallel}=-2.16$ MHz is the longitudinal component of the hyperfine tensor describing the NV electron and the $^{14}$N nuclear spin interaction.
From the Lorentzian fit to this central line we extract a resonance linewidth of $\approx$ 0.9~MHz, close to what is measured at smaller ($\approx$~50 G) magnetic fields, highlighting the negligibly small inhomogeneities in the B field (see section II of the supplementary materials \cite{supp} for more technical details). 

The other features correspond to the two other $^{14}$N nuclear spin–conserving electronic transitions as well as the four electronic transitions related to hyperfine coupling with the nuclear spin of the A, B, C, and D $^{13}$C families \cite{Dreau2012,Smeltzer_njp}. Proximal $^{13}$C nuclei from family A have already been observed in ODMR with a resolution of about 2~MHz (see, e.g., Refs. \cite{Jarmola,Fischer}).
The frequencies of the four $^{13}$C–NV transitions we observe can be approximated by $\nu \pm |A_{\parallel} \pm \gamma_n B|$ far from level crossings in the ground state (the $\pm$ signs depend on the signs of the hyperfine tensor $A_{\parallel}$ for each family).
These analytical expressions provide a good estimate of the peak positions. In this measurement, the precision with which one can extract the $^{13}$C–NV hyperfine tensor from ODMR spectra is nevertheless limited by $1/T_2^*$, where $T_2^* \approx 1 \mu $s is the electron spin coherence time. Improving beyond this resolution is inherently challenging with highly doped NV ensembles. 

To enhance spectral resolution, a common approach is to employ {Electron Spin Echo Envelope Modulation (ESEEM)} \cite{Schweiger}. The sequence is illustrated in the inset of Fig.~\ref{ODMR}-b). Figure~\ref{ODMR}-b) displays the photoluminescence (PL) as a function of the free precession time, after subtracting the slowly varying spin-echo envelope, which decays with a characteristic time of $T_2 \approx 7(3),\mu$s (see Section~III of the Supplementary Materials \cite{supp}). The data reveal fast oscillations at a frequency of 521(5)~kHz (see the Fourier transform below), corresponding to the $^{13}$C Larmor precession at a magnetic field of $487(1)$~G. This is clear evidence that the electron spin experiences a periodically modulated hyperfine field from coupled $^{13}$C nuclei.
Applying a transverse magnetic field could, in principle, allow one to distinguish the contributions from the different $^{13}$C families in the Fourier spectrum and extract the complete hyperfine tensors \cite{Childress2006}. However, the achievable resolution would be limited to $1/T_2 \approx 140$~kHz. Moreover, inequivalent $^{13}$C sites within the same family would precess at slightly different frequencies, leading to a broadening of the Fourier peaks.

In the rest of the manuscript, we will use another method that is closer in spirit to NMR to measure nuclear spins and hyperfine tensor components with higher spectral resolution, enabling precise estimate of the full $^{13}$C-NV hyperfine tensor.
When a green laser is continuously applied to the NV centers, the fraction of the electronic spin population in a given electronic spin state will flip depending on the nuclear spin state. 
Applying an RF tone in the NV ground state to flip a polarized nuclear spin when at the ESLAC 
will indeed result in an electron-nuclear spin flip-flop in the optically excited state. The electron flip to the $\ket{-1_e}^{\rm ex}$ state, will provoke a non-radiative decay to the NV metastable state, thus lowering the PL rate. In turn, the NV PL thus depends on the nuclear spin state \cite{smeltzer_pra, Steiner}. The protocol will hereafter be 
called ODNMR \cite{Jarmola_1}. As we shall see, detecting $^{13}$C-NV couplings in the $\ket{\pm 1}$ states {\it as well as} in the $\ket{0_e}$ electronic states enables precise determination of the full hyperfine tensor. 

 \begin{figure}
\includegraphics[width=9cm]{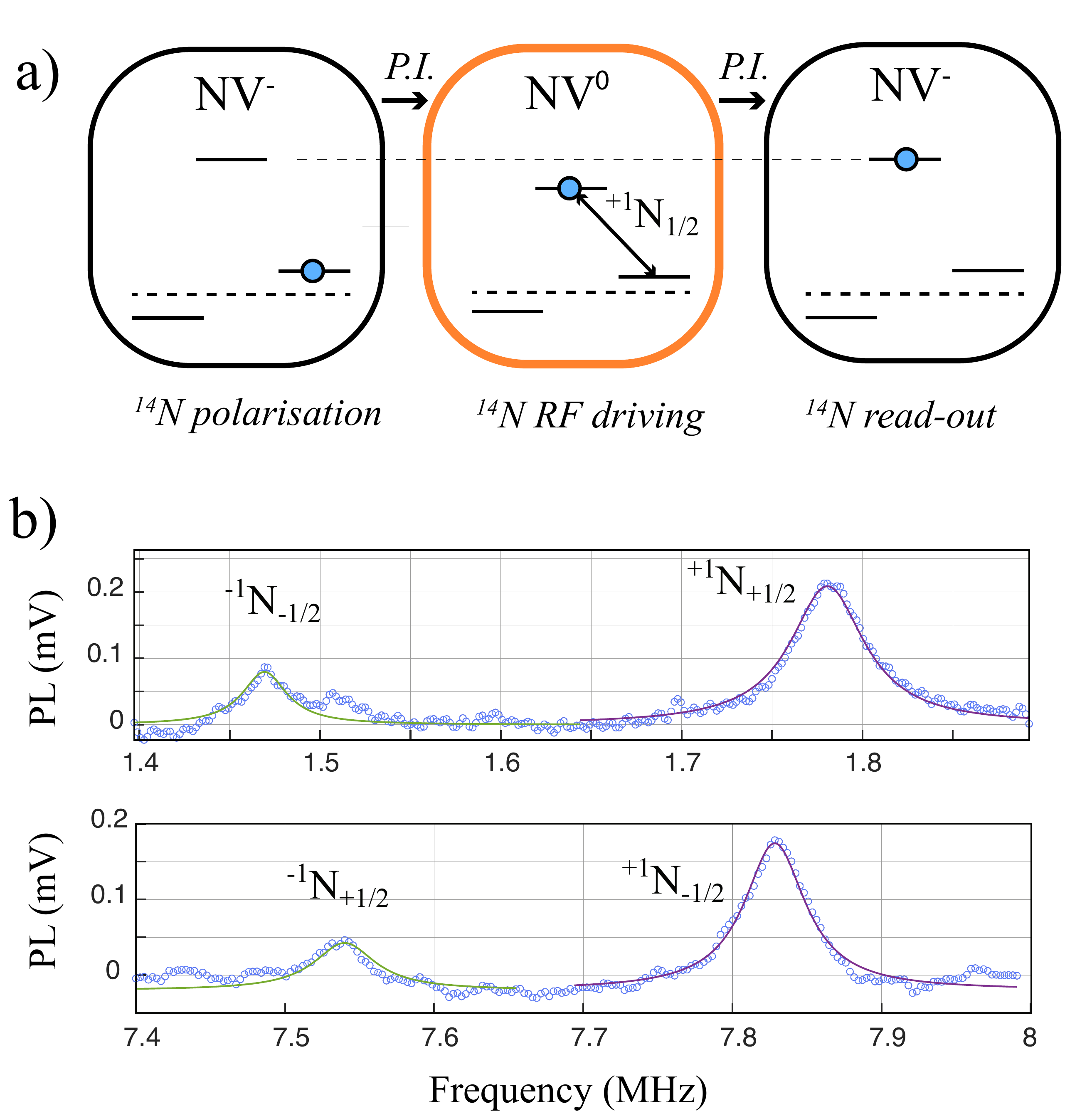}
\caption{a) Suggested process enabling NMR in the NV$^0$ charge state : Photo-ionization (P.I.) followed by radio-frequency excitation of the $^{14}\mathrm{N}$ spin in the NV$^0$ charge state and read-out in the NV$^-$ state. b) Optically Detected Nuclear Magnetic Resonance (ODNMR) spectrum at B= 486.8~G showing resonances associated with $^{14}$N nuclear spins coupled to the NV$^0$ center in the $\ket{\pm/1/2_e}$ manifolds.  }
\label{ODNMRnv0}
\end{figure}
\section{$^{14}$N NMR in the NV$^-$ state}

In this part, we study ODMNR under 10~mW of continuous-wave laser power. 
At this power level, the laser does not fully polarize the NV center in the $\ket{0_e}$ state so that nuclear spin transitions in the electronic states $\ket{\pm 1_e}$ can be detected without applying microwave \cite{smeltzer_pra} (See SM, section IV).

Fig. \ref{ODNMR}-a) shows the NV PL as a function of RF drive frequency up to 14.4 MHz, featuring many sharp peaks. The first one, indicated by a downward pointing arrow, is close to
$\gamma_n B=521$ kHz under a magnetic field of 487~G.
Thus, this peak may be attributed to the $^{13}$C NMR transitions in the $\ket{0_e}$ state. This will be confirmed in the next section. The other most pronounced peaks {have already been reported} \cite{Jarmola_1, lourette2023}, and are associated to $^{14}$N spins in the three NV$^-$ center electronic states. 

Fig. \ref{ODNMR}-b) and c) show a zoom on the four nuclear spin transitions in the \(\ket{\pm 1_e} \) manifolds of the NV$^-$ center. The low employed optical power level permits observing all of the four transitions with high sensitivity, without requiring microwave pulses to drive the electron to the $\ket{\pm 1_e}$. Notably, the $^{-1} N_{-1}$ and $^{+1}N_{+1}$ transitions are both much lower than the $^{-1} N_{-1}$ transition and almost have the same contrast, although $\ket{+1_N}$ state is better polarized. This phenomenon is likely to be caused by the ODNMR read-out protocol. At the ESLAC, the flip-flop only involves the $\ket{-1_e}$ and $\ket{0_e}$ states, so only nuclear spins in the $\ket{-1_e}$ are read-out efficiently. {This effect boosts the read-out efficiency on the $^{-1} N_{-1}$ NMR transition and lowers it on the $^{+1} N_{+1}$ transition \cite{Jarmola_1}. }

Using this spectrum, full information about the $^{14}$N-NV$^-$ hyperfine tensor can be obtained (see Supplementary materials \cite{supp}, section VIII).

\section{$^{14}$N NMR in the NV$^0$ state}

We now turn to observations of ODNMR in the {\it neutral charged state} of the NV center, written NV\textsuperscript{0}.
Not visible on Fig. \ref{ODNMR}-a) are pairs of peaks around 2 and 8 MHz. Fig.~\ref{ODNMRnv0}-b) shows two scans around these transitions. 
These are attributed to the driving of the nuclear spin of the nitrogen 14 atoms in the NV\textsuperscript{0} state.
In the NV\textsuperscript{0} state, the electron spin of the defect can be \( S = 1/2 \) or \( S = 3/2 \) \cite{Felton_neutral, Razinkovas}.

\begin{table}[h]
    \centering
    \renewcommand{\arraystretch}{1.2}
\begin{tabular}{lc|lc}
\hline\hline
\multicolumn{2}{c|}{NV$^0$ ODNMR resonances} & \multicolumn{2}{c}{$^{14}$N Nuclear spin parameters} \\
\hline
Label & Frequency (MHz) & Param. in NV$^0$  & Value (MHz) \\
\hline
$^{+1}$N$_{-1/2}$ & 7.828(3) & $|Q_0|$         & 4.655(5) \\
$^{-1}$N$_{-1/2}$ & 1.472(9) & $A^0_\parallel$ & 6.057(3) \\
$^{+1}$N$_{+1/2}$ & 1.782(3) & $A^0_\perp$     & 3.9(6) \\
$^{-1}$N$_{+1/2}$ & 7.540(8) &               &          \\
\hline\hline
\end{tabular}
\caption{Measured ODNMR transition frequencies (left) and $^{14}$N nuclear spin parameters (right) in the NV$^0$ state.}
\label{NVnot}
\end{table}

Neglecting second-order electron-nucleus flip-flop processes for now, the effective spin Hamiltonian for the $^{14}$N nucleus in the $S=1/2$ state of the NV\textsuperscript{0} center reads:
\begin{align}\label{NVOcal}
\mathcal{H}_N^{\text{NV}^0}/\hbar = Q_0 I_z^2 +\gamma_n^{(N)} B_z I_z + m_s A^0_{\parallel}  I_z,
\end{align}
where \( Q_0 \approx -4.655~\,\mathrm{MHz} \)  \cite{control_charge_state, Waldherr} is the nuclear quadrupolar interaction, and \( A^0_{\parallel}\approx \rm 6.06~MHz \) is the hyperfine coupling constant between the \( S = 1/2 \) electron spin and the nuclear spin \cite{Waldherr}.  

In contrast to the NV\textsuperscript{-}, the NV\textsuperscript{0} state lacks spin-selective optical transitions, which restricts its applicability in coherent spin-control schemes. As illustrated in Fig.~\ref{ODNMRnv0}-a), however, photo-ionization (P.I.) may allow readout of the NV\textsuperscript{0} nuclear spin states, provided that P.I. does not disturb the nuclear spins. In this protocol, $^{14}$N nuclear spin polarization is first established in the NV\textsuperscript{-} charge state. A subsequent two-photon process \cite{Siyushev} converts NV\textsuperscript{-} into NV\textsuperscript{0}, where both the hyperfine interaction and quadrupole moment differ. Radio-frequency driving can then flip the $^{14}$N spin, after which a second P.I. process restores the NV to the negatively charged state, enabling optical readout of the $^{14}$N nuclear spin.

\begin{figure*}
\includegraphics[width=17cm]{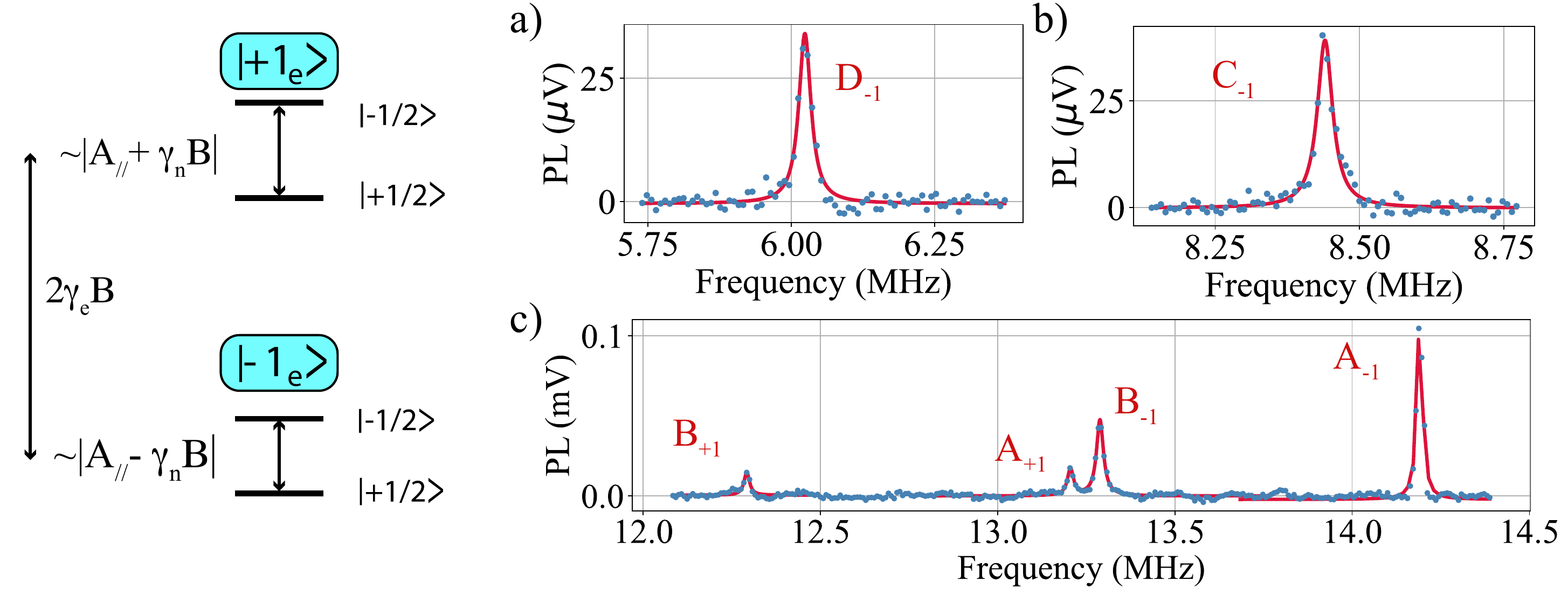}
\caption{(Left) Energy levels of the $^{13}$C nuclear spins in the $\ket{\pm 1_e}$ states. a),b) and c) are close-up of the ODNMR resonances arising from the coupling between $^{13}$C nuclei and NV electronic spins in the $\ket{\pm 1_e}$ magnetic states. Blue dots correspond to experimental data and red lines to Lorentzian fits. The resonance frequencies are reported on Tab.\ref{tab:ms_splitting}. Resonances are labeled by the $^{13}$C family and using the NV electronic spin in index.}
\label{C13_msm1}
\label{tab:odnmr_full}
\end{figure*}

The spectral positions of the four peaks in Fig.~\ref{ODNMRnv0}-b) are reported in the table \ref{tab:odnmr_full}.
There is good agreement between the predicted positions of the peaks and the expectations found from Eq.~(\ref{NVOcal}).
{There is also good agreement with the reported values \cite{Waldherr}.}
Interestingly, the contrast of the four $^{14}$N nuclear spin transitions in the NV$^0$ state is fully consistent with a polarization of the nucleus in the 
$\ket{1_N}$ state. This shows that a large $^{14}$N nuclear spin polarization is preserved after photo-ionization.
Also, the P.I. transfer enables unambiguous CW read-out of the spin-polarization. 
This is to be compared with the read-out of the $^{14}$N nuclear spin in NV$^-$ where ODNMR biases the read-out contrast, as discussed in Section II.

We now seek to provide quantitative estimates of the NV$^0$ hyperfine and quadrupole tensors. To this end, we employ the formulae in the Supplementary Materials \cite{supp} (Section VIII), which include non-secular contributions.
The values are reported in the table \ref{NVnot}. 
We also extract the $A^0_\perp$ component, which has not been reported thus far to the best of our knowledge. Note that our method involves a subtraction of two frequencies that are very close to each other, so the uncertainty in $A^0_\perp$ is larger than in the estimation of $Q_0$ and $A^0_\parallel$.

\section{$^{13}$C NMR}

\subsection{ODNMR in the NV magnetic states}

Let us now turn to additional peaks that are not visible in Fig.~\ref{ODNMR}-a), which are displayed in Fig.~\ref{C13_msm1}. 
The central frequencies of these peaks, obtained by fitting Lorentzian functions, are reported in Table~\ref{tab:ms_splitting} (left). 
The peaks labeled A$_{m_s}$, B$_{m_s}$, and D$_{m_s}$ are in excellent agreement with the expected transitions in the $\ket{\pm 1_e}$ NV electronic states for the three families of $^{13}$C spins, as previously observed by Smeltzer {\it et al.} \cite{Smeltzer_njp} using pulsed ODMR with interleaved RF drive in the $\ket{\pm 1_e}$ manifolds. {The position of each of the $^{13}$C atoms with respect to the NV center is depicted in Fig. 1.}
The previously unreported peak at 8.45~MHz observed here can be attributed to C$_{-1}$, based on the ODMR results from Dr\'eau {\it et al.} \cite{Dreau2012}.

Interestingly, sharp NMR peaks are observed, with a width on the order of the one measured in the single spin experiments \cite{Smeltzer_njp}, but here using many NV centers coupled to $^{13}$C atoms at different sites, thus improving NMR sensitivity. All the sites  -- whose Hamiltonian differ one by $\phi$ (see Eq.~(\ref{Ham2}))--  for a given family are indeed degenerate in the $\ket{\pm 1_e}$ states.

Diagonalizing the Hamiltonian Eq.~(\ref{Ham2}) using the values of $A_{\parallel}$ from 
\cite{Smeltzer_njp, Dreau2012} and neglecting $A_{\perp}$ already yields a fair agreement with the observed transition frequency.
We postpone the estimation of $A_{\parallel}$ to when $A_{\perp}$ will be extracted. The latter can indeed be estimated efficiently, independently on $A_{\parallel}$, in the $\ket{0_e}$ state.

\begin{table}[htbp]
\centering
\renewcommand{\arraystretch}{1.2}

\begin{tabular}{l c c c c c}
\toprule
\multicolumn{6}{c}{$^{13}$C ODNMR resonances in NV$^{-}$ center} \\
\midrule
& \multicolumn{2}{c}{$\ket{+1_e}$} & \multicolumn{2}{c}{$\ket{-1_e}$} & $\ket{0_e}$ \\
\cmidrule(lr){2-3} \cmidrule(lr){4-5} \cmidrule(lr){6-6}
Label & Freq. (MHz) & Label & Freq. (MHz) & Label & Freq. (kHz) \\
\midrule
$A_{+1}$ & 13.2052(9) & $A_{-1}$ & 14.1907(4) & $A_0$ & 469.79(3) \\
$B_{+1}$ & 12.292(1)  & $B_{-1}$ & 13.2985(3) & $B_0$ & 480.27(5) \\
          &            & $C_{-1}$ & 8.4402(5)  & $C_0$ & 499.29(9) \\
          &            & $D_{-1}$ & 6.0235(5)  & $D_0$ & 516.58(5) \\
\bottomrule
\end{tabular}

\caption{$^{13}$C ODNMR resonances in the three electronic spin states $\ket{\pm 1_e}$ and $\ket{0_e}$ of the NV$^{-}$ center at $B = 486.8~\text{G}$ aligned with the NV axis. Values and uncertainties are obtained from Lorentzian fits.}
\label{tab:ms_splitting}
\end{table}

\subsection{ODNMR and coherent control in the $\ket{0_e}$ state}

We now study ODNMR in the $\ket{0_e}$ state, where to the best of our knowledge, no $^{13}$C spins have yet been detected. 
Since the $\ket{0_e}$ state is non-magnetic, the NMR peaks for all $^{13}$C families are expected to lie at around $\gamma_n {B}\approx 520$ kHz, with B= 486.8 G.
 This, in fact, almost coincides with the first broad peak observed in Fig. \ref{ODNMR}. 
 
\subsubsection{ODNMR in the $\ket{0_e}$ state -- theory}\label{theorysec}

\begin{figure}
\includegraphics[width=8cm]{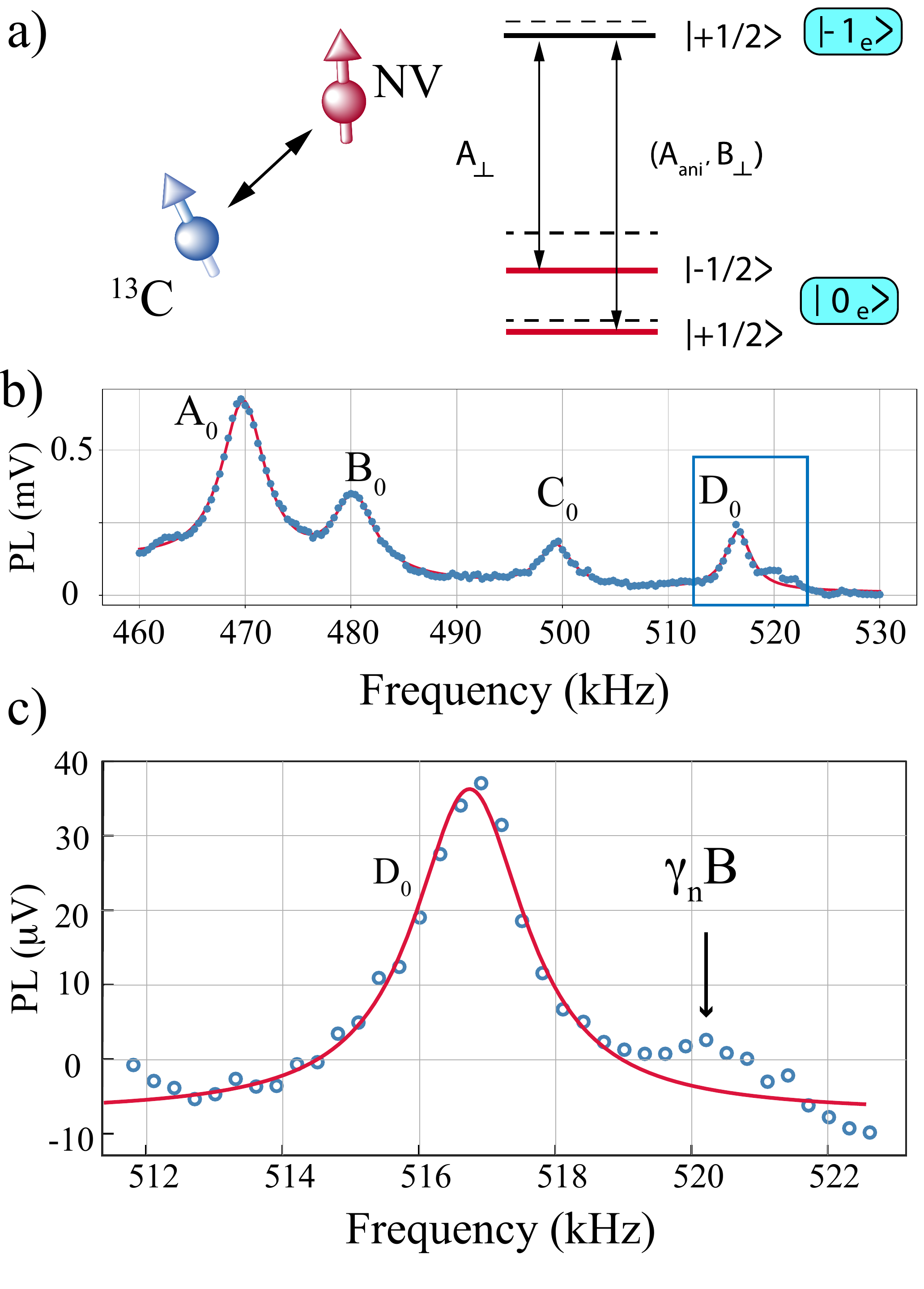}
\caption{
a) Coupled $^{13}$C - NV spins and simplified level structure ($\ket{+1_e}$ state not depicted) showing the mechanism by which the two quasi-degenerate $^{13}$C states are coupled in the $\ket{0_e}$ state. 
b) Optically detected nuclear magnetic resonances (ODNMR) of $^{13}$C nuclei coupled to the NV center electronic spin in the $\ket{0_e}$ state. The magnetic field $\mathbf{B}$ is aligned along one NV axis. B = 488.15 G. Four resonances are fitted with Lorentzian profiles, the  parameters are reported in Tab.\ref{tab:deltaE}. c) Enlarged scan around the peak $D$, showing an extra peak, possibility related to all other weakly coupled $^{13}$C families. }
\label{ms=0}
\end{figure}

Figure \ref{ms=0}-b) displays an ODNMR scan acquired with 30 dB lower RF power than in the data shown in Fig.~\ref{ODNMR}, revealing four narrow peaks in the vicinity of $\nu=\gamma_n B$ now. To account for the origin of these peaks, one must consider second-order hyperfine interactions with the $\ket{\pm 1_e}$ NV spin states. Although the NV is nominally in a non-magnetic state, the transverse component of the hyperfine interaction can induce an effective electronic magnetic moment, which in turn shifts the nuclear energy quasi-degenerate levels \cite{petta2016, chen_prb, Childress2006}. 

Here, we derive an effective two-level Hamiltonian for a single 
$^{13}$C spin in the subspace $\ket{0_e}$ taking into account the hyperfine interaction in all NV states. 
Using a second order perturbation theory of degenerate states (see supplementary method \cite{supp}, section VI and VII) we obtain 
$$
\begin{aligned}
\frac{H_{\rm eff}^{\ket{0_e}}}{\hbar} =  \frac{1}{2}\begin{pmatrix}
{}\gamma_nB_z + \nu & {}\gamma_nB_x+h_\perp\\
{}\gamma_nB_x+h_\perp^*&-{}\gamma_nB_z - \nu \\
\end{pmatrix},
\end{aligned}
$$
in the basis $\ket{\pm 1/2}$, 
with: 

\bea\label{hperpnu}
    h_\perp &=&  -\frac{\gamma_e A_\perp}{\Delta^2}  (A_{\rm ani}e^{i\phi} B_z + 2 D B_x )  \\
    \nu &=& - \frac{\gamma_e}{\Delta^2}(2D \cos\phi A_{\rm ani} B_x+ A_\perp^2 B_z).
\eea
We wrote $\Delta^2=D^2-(\gamma_eB_z)^2$.
The diagonal terms are due to virtual transitions through states $\ket{1_e}$ and $\ket{-1_e}$ by the combined action of $A_{\perp}$ and $A_{\rm ani}$. This process is shown in Fig. \ref{ms=0} (note that the state $\ket{+1_e}$ is not depicted, but it also contributes to the frequency shifts). They involve flip-flop processes between the electronic and nuclear spins followed by electronic spin flips which preserve the nuclear spin. 
The Hamiltonian can in fact be recast in terms of an effective Zeeman Hamiltonian with a $g$-tensor capturing the effects of the anisotropic and transverse components of the hyperfine tensor (see \cite{Childress2006, Duarte} and the SM, section IX). 

From this Hamiltonian we can compute the shift in energy between the two  $\ket{\pm 1/2}$ eigen-states. It is given by: 
\bea\label{shift0}
\Delta E =\hbar  \sqrt{({}\gamma_nB_z+\nu)^2+|{}\gamma_nB_x+h_\perp|^2}.
\eea

In the particular case where $B_x=0$ we get: 
\bea\label{shiftE} \nonumber\Delta E = |{}\gamma_nB_z|\sqrt{\bigg(1+\frac{\gamma_e}{{}\gamma_n}\frac{A_\perp^2}{\Delta^2}\bigg)^2 + \Big(\frac{\gamma_e}{{}\gamma_n} \frac{2A_\perp A_{\rm ani}}{\Delta^2}\Big)^2}.\\
\eea

We observe that the energy splittings, and consequently the RF resonance frequencies, differ for each $^{13}$C family, due to variations in $A_\perp$ and $A_{\rm ani}$.

\begin{figure}
\includegraphics[width=8cm]{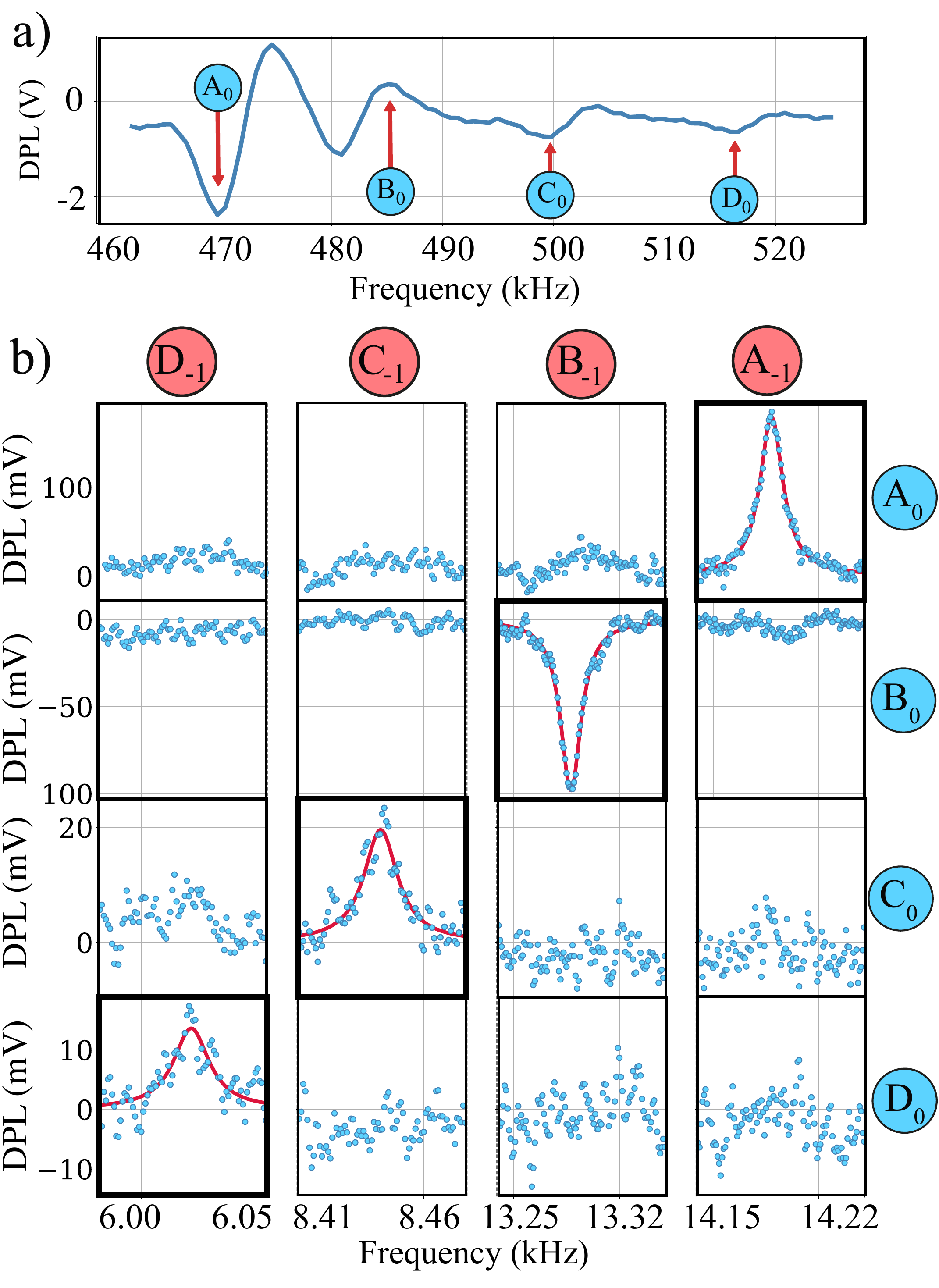}
\caption{a) ODNMR recorded from 460~kHz to 525~kHz, using a lock-in detection. DPL = demodulated photoluminescence. The FM modulation frequency of the RF field is 2~kHz. Frequencies shown with arrows are the one chosen to realize the two-tones spectroscopy shown below. b) DPL scans around $^{13}$C transitions in the $\ket{-1_e}$ states (indicated by the red circles on top) with a second modulated RF near one of the four $^{13}$C families in the $\ket{0_e}$ manifold (indicated by blue circles on the right). Red continuous curves are Lorentzian fits to the data.    
}
\label{lock}
\end{figure}

\subsubsection{Two-tones ODNMR spectroscopy}

\begin{figure*}
\includegraphics[width=13cm]{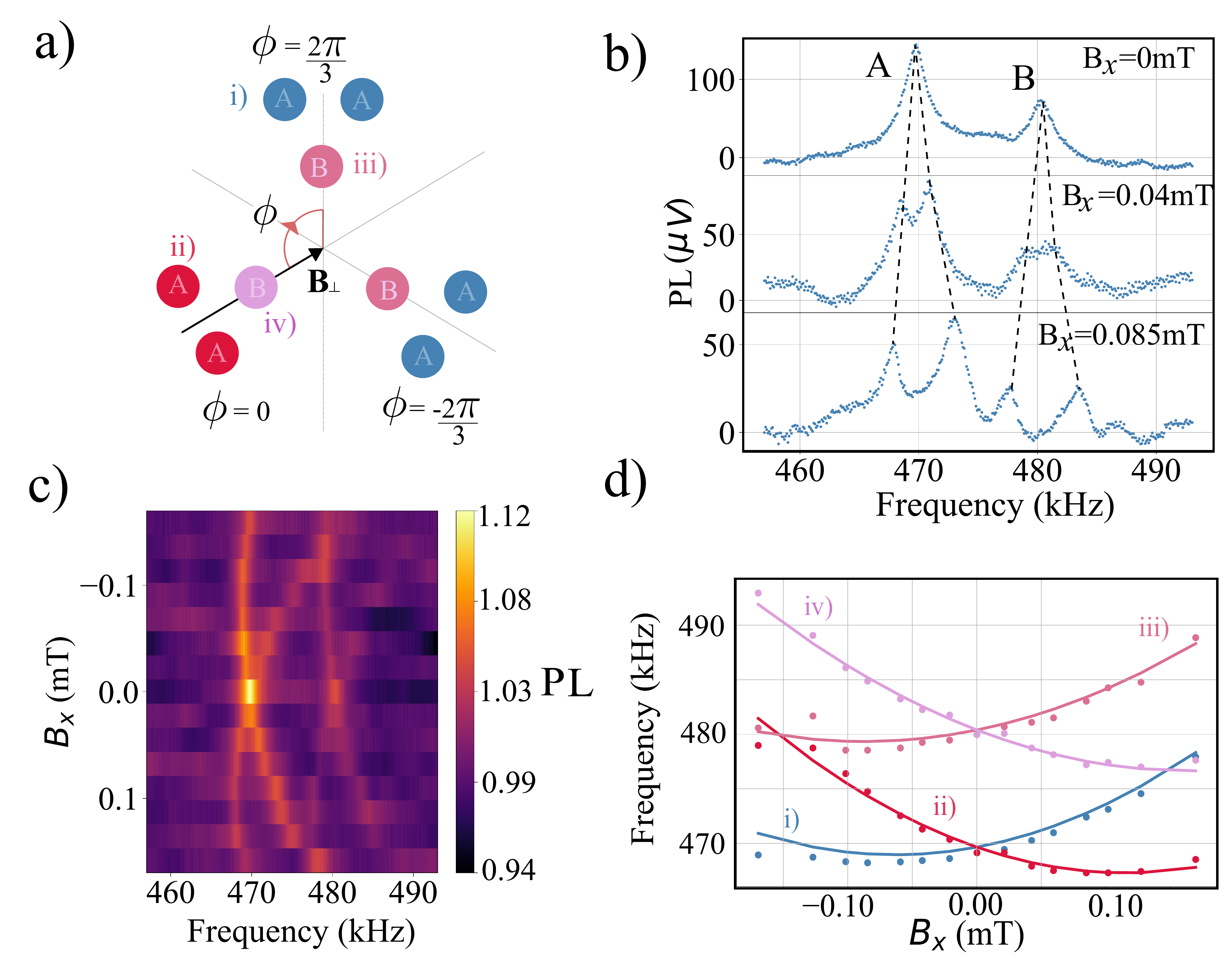}
\caption{
a) Schematic representing the $^{13}$C repartition, of families A and B around an NV center (not shown). The direction of ${\bm B}_\perp$ shown here is the one employed in the measurements shown in b)-c) and d).
b) ODNMR spectra for families A and B for the three values of $B_x$ shown on the top right (here $\phi=0$ so ${\bm B}_\perp = B_x \bm e_x$). 
Dashed lines indicate the evolution of the peaks' maxima obtained using Lorentzian fits. c) 2D plot showing the photoluminescence as a function of the transverse magnetic field amplitude and RF frequency for families A and B. 
d) Transition frequencies as a function of the transverse magnetic field $B_x$. The data points (dots) are extracted from the maxima in the ODNMR spectra of Fig.\ref{ms=0}-b). Continuous lines correspond to the fit using Eq. \ref{shift0}, where $\phi$ is set to 0 for traces i) and iii) and $2\pi/3$) for traces ii) and iv). $A_{\rm ani}$ is used as a free parameter in these fits. Each trace is associated to a $^{13}$C as labeled in a).}
\label{effet_bx}
\end{figure*}

The values of $A_\perp$ have at present not been measured for the four families A, B, C and D.
We may tentatively attribute the resonances in the $\ket{0_e}$ manifold to families A$_0$, B$_0$, C$_0$ and D$_0$ in ascending order of transition frequencies, following the $A_{\parallel}$ trend. The extra smaller peak that is seen at $\gamma_n B\approx 520~G$ (a close-up is shown in Fig.~\ref{ms=0}-c)) would then likely correspond to the other weakly coupled $^{13}$C families, including possible contributions from diffusive $^{13}$C spins that are not coupled to the NV center directly. 

This interpretation is not entirely satisfactory however. Theories suggest that the $A_\perp$ ordering between families can differ significantly from that of $A_{\parallel}$ \cite{Gali, PhysRevB.92.115206}. 
To resolve this ambiguity, we perform two-tones correlation spectroscopy involving the $\ket{-1_e}$ electronic states in which each $^{13}$C family was previously unambiguously identified (see Section IV-A).
The idea is to record demodulated PL lock-in  signals (DPL) by scanning an RF tone close to resonances in the $\ket{-1_e}$ manifold while a second modulated RF tone is on resonance with one of the peaks in the $\ket{0_e}$ state. 
First, we perform a lock-in detection close to the A$_0$, B$_0$, C$_0$ and D$_0$ peaks with a frequency-modulated RF signal. 
Fig.\ref{lock}-a) shows the corresponding demodulated photoluminescence (DPL) scan, where all peaks seen in Fig. \ref{ms=0}-b) are clearly identified.  
Next, as indicated by the red arrows, we park the modulated RF signal around one of the four transitions in this $\ket{0_e}$ manifold. We then detect a DPL around the four A,B,C, and D transitions in the $\ket{\pm 1_e}$ state for four different pump frequencies using a second RF generator.

The principle of this technique is that a demodulated detection scan around one of the four nuclear transitions—for instance B$_{-1}$—yields a signal only if a modulated pump is applied on B$_0$, thereby depleting the $\ket{0_e}$ state. Since the lock-in detection is sensitive exclusively to resonances involving $^{13}$C nuclei from the same spin family, a DPL signal should thus appear solely for the B$_{-1}$–B$_0$ combination.

Figure \ref{lock}-a) presents the results of sixteen measurements obtained with the two-tones correlation spectroscopy technique. A signal within a given family appears only when the pump is tuned to that same spin family, with no cross-signals observed. This allows us to unambiguously identify the families in the $\ket{0_e}$ subspace and confidently assign the four resonances in Fig.~\ref{ms=0}-b) to the A, B, C, and D families, in ascending order.

One essential feature in the theory presented in the section \ref{theorysec}, is that the central position of the A$_0$, B$_0$, C$_0$ and D$_0$ peaks depend solely on $|A_\perp|$ to lowest order. The third term under the square root in Eq.~(\ref{shiftE}) is indeed negligibly small as we shall see. This fact greatly facilitates the reconstruction of the hyperfine tensor.
From the central position of these peaks, we can thus extract the absolute value of $A_\perp$ independent on $A_{\parallel}$. 
This is reported in table \ref{tab:deltaE}, setting the sign of $A_\perp$ to be the same as the sign of $A_{\parallel}$ (see section VIII of the SM).

Once the $A_\perp$ are extracted, another identification check can be done by comparing the amplitude of each of the four peaks. 
Their amplitudes are proportional to $p\times \Omega^2 \times N_{\rm site}$ where $p$ is the polarization of the $^{13}$C spin, $\Omega$ the RF Rabi frequency and $N_{\rm site}$, the number of sites. 
One can use the formulas in \cite{chen_prb} to estimate $\Omega=\gamma_n' B_{\rm RF}$ where $\gamma_n'$ is the hyperfine enhanced gyromagnetic ratio --related to $A_\perp$-- and $B_{\rm RF}$ is the RF magnetic field amplitude. Using the polarization $p$ measured in \cite{Dreau2012}, we find that the amplitude ratios between each peak match very well this prediction (see supplementary materials  \cite{supp}, section IX-B).

\subsubsection{Lifting site degeneracy}

Theory also predicts that there is a dependency of the frequency of the various sites {\it within the same family} (see Fig. 1-a)) to a transverse magnetic field. As is manifest in Eq.~\ref{shift0}, when both $\phi$, namely the azimuthal angle that identifies a site in the same family, and $B_x$ are included in the theory, the energy levels depend on $\phi$. $^{13}$C nuclear spins states of the same family are then no longer degenerate in a transverse B field. $A_{\rm ani}$ then contributes at leading order to the shift, allowing specific spatial $^{13}$C–NV configurations to be distinguished spectrally. 
In order to show this effect while still retain sufficient signal to noise, we performed an experiment where the transverse component of the magnetic field is set as sketched in Fig. \ref{effet_bx}-a).
The magnetic field direction is set in the middle of two A carbon 13 atoms so that only two B field projections onto the $^{13}$C-NV axes differ, instead of six for the family A and three for the family B. 
A clear splitting of the A and B transitions due to the transverse component of the magnetic field is observed in \ref{effet_bx}-b) as it is stepped up to 0.085 mT.  Fig. \ref{effet_bx}-c) shows a scan of the transition frequencies in families A and B, when the transverse component of the B field spans -0.15 mT to 0.15 mT. The evolution of the transition frequencies as a function of $B_x$ is shown in Fig. \ref{effet_bx}-d). Each of these four curves are fitted by Lorentzians, thus providing a precise and uni-valued measurement of $A_{\rm ani}$.
{The hyperfine components, including the $A_{\parallel}$ components obtained from the transitions frequencies in the table \ref{tab:ms_splitting}, are recorded in the table \ref{tab:deltaE} and are compared to the theory performed in 
\cite{Ivady}. A good agreement between the theory and the experiment is obtained, highlighting the accuracy of the ODNMR and of the theoretical methods used in \cite{Ivady}.}

\begin{table}[h]
    \centering
    {
    \begin{tabular}{|c|c|c|c|c|c|c|c|c|c|c|c}
        \hline \hline
        $^{13}$C Family  & \multicolumn{2}{|c|}{$A_\perp$ (MHz)} & \multicolumn{2}{|c|}{$A_{\rm ani}$ (MHz)} & \multicolumn{2}{|c|}{$A_\parallel$ (MHz)}\\
        \cline{2-7}
        (\# sites) & Exp. &Theo.& Exp. &Theo.& Exp. &Theo.\\ 
        \hline
         A (6) & 15.6(1) &15.54  & 1.56(8)&1.62 & 13.7496(1) &13.38\\
        \hline
        B (3) & 14.00(7)&13.88 & 1.8(1) &1.88 & 12.795(2) &12.62 \\
        \hline
        C (3) & -10.4(1)&-10.17& 0.9(2)&0.81  & -8.9(2)&-8.84    \\
        \hline
        D (6)& -5.2(2)&-5.64 &     - &0.90   & -6.51(2)&-7.00   \\

        \hline
    \end{tabular}
    }
    \caption{{
    Values of the $^{13}$C-NV hyperfine interaction terms for the families A, B, C and D from the experiment and from the theory developped in \cite{Ivady}. The $A_\perp$ components are evaluated using data from Fig.~\ref{ms=0}-a). The $A_{\rm ani}$ components are extracted from Fig.~\ref{effet_bx}-a) using a fit with Eq.~(\ref{shift0}). $A_{\parallel}$ are extracted from the transitions listed in table \ref{tab:ms_splitting}.}
   }
    \label{tab:deltaE}
\end{table} 

\subsubsection{Coherent control of $^{13}$C nuclear spins}

To use the $^{13}\mathrm{C}$ nuclear spins for applications, achieving quantum coherent control of their spin states is essential.  
Here, we show coherent control of $^{13}$C nuclear spin ensembles in the $\ket{0_e}$ manifold.

As a preliminary step, we demonstrate Ramsey oscillations on the $\ket{+1_N} \rightarrow \ket{0_N}$ transition of the $^{14}\mathrm{N}$ nuclear spin in the $\ket{0_e}$ electronic state.  
The duration of a $\pi/2$ pulse is first calibrated via Rabi oscillation measurements.  
We then perform a standard Ramsey sequence using a radio-frequency tone detuned by approximately 8\,kHz from the central transition frequency at 5.091\,MHz. 
\begin{figure}
\includegraphics[width=8cm]{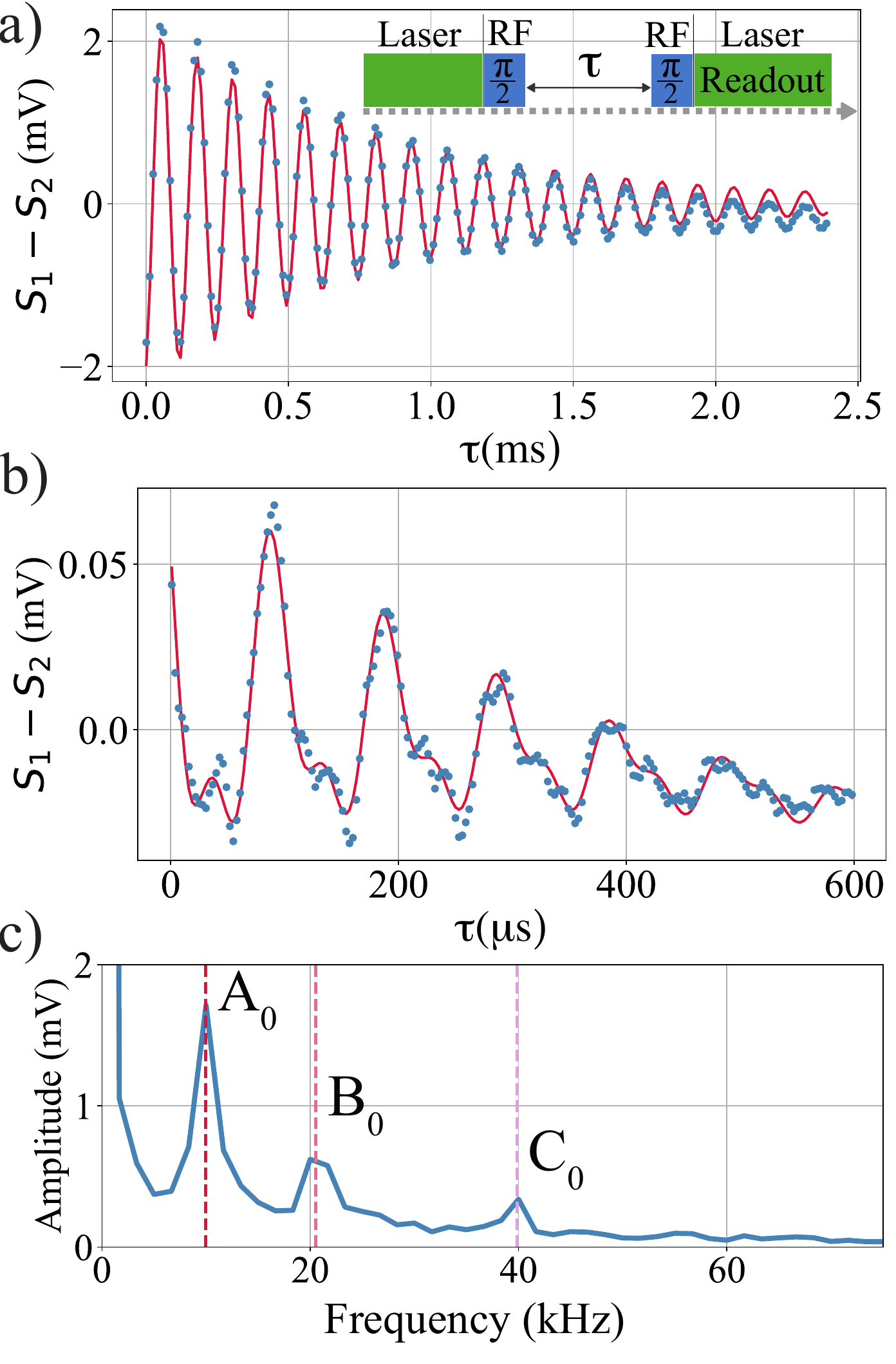}
\caption{a) Ramsey oscillation of the $^{14}$N nuclear spin state at a frequency detuning of 8 kHz. The sequence is detailed in the top right of the figure. b) Ramsey oscillation of the $^{13}$C nuclear spin. The RF is fixed at 460~kHz. Experimental data are fitted with a sum of two sinus with exponential decay. c) Fourier transform of the Ramsey oscillation signal. Three resonances corresponding to families A,B and C in the $\ket{0_e}$ state are visible.}
\label{ramsey}
\end{figure}
Figure~\ref{ramsey}-a) displays the resulting decaying Ramsey oscillations.  
A fit with an exponentially decaying sinusoid yields a detuning of 8\,kHz and a dephasing time $T_2^*$ of approximately 864(7)\,$\mu$s.  
This value is close to the $T_1$ time of the NV electronic spins (see Supplementary materials, \cite{supp} section V).  
It is also close to the decay time estimate that one can derive from the electronic $T_2^*$ and ratio of the nuclear to electronic gyromagnetic ratios.

We now move on to performing coherent control of the $^{13}$C ensemble in the $\ket{0_e}$ state. 
One difference with the nitrogen nucleus is the non-degeneracy of the A, B, C and D peaks, even in the $\ket{0_e}$ state (see Fig.~\ref{ms=0}).
Fig. \ref{ramsey}-b) shows a Ramsey measurement with an RF tone set at 460~kHz.
We fit this data by a sum of two sinusoids with frequencies $\omega_A=$ 9.87(3)~kHz and $\omega_B$=20.34(6)~kHz and a single exponential decay.
 The Fourier transform of the signal in Fig. \ref{ramsey}-b) features these two frequencies, plus another one around 40~kHz from family C. From the fit we extract a decay time for the $^{13}$C nuclear spins coherence time $T_2^*\approx$~203(9)~$\mu s$. The ratio of the decoherence time of $^{13}$C and $^{14}$N nuclear spins is of the same order of magnitude as the ratio of their gyromagnetic ratios indicating that the decoherence times have the same origin in both species.

We conclude this section by noting that, in principle, all-optical pulsed detection of $^{13}$C nuclear spins can be performed in a similar manner to that of the NV center’s $^{15}$N nucleus, as demonstrated in Ref.~\cite{All_optical} (both having a spin 1/2), albeit with a smaller contrast and with multiple frequency components instead of just one.

\subsection{Coupled pairs of $^{13}$C spins}
\begin{table}[htbp]
\centering
\renewcommand{\arraystretch}{1.2}
\setlength{\tabcolsep}{10pt}

\begin{tabular}{l c c}
\toprule
\multicolumn{3}{c}{\textbf{$^{13}$C spin-pair transition frequencies (kHz)}} \\
\midrule
Pair & Theory & Experiment \\
\midrule
A--A & 359.2(1) & 360.1(3) \\
B--B & 391.2(1) & 392.0(6) \\
C--C & 455.5(1) & 452.2(4) \\
D--D & 504.7(1) & {} \\
\addlinespace
A--B & 375.1(1) & 376.2(6) \\
A--C & 409.0(1) & 412.9(3) \\
A--D & 449.3(1) & 452.2(4) \\
B--C & 422.5(1) & 426.8(2) \\
B--D & 460.4(1) & 462.1(7) \\
\bottomrule
\end{tabular}

\caption{Theoretical predictions and experimental observations for $^{13}$C spin-pair transition frequencies (in kHz).}
\label{coupledtable}
\end{table}

\begin{figure}
\includegraphics[width=8cm]{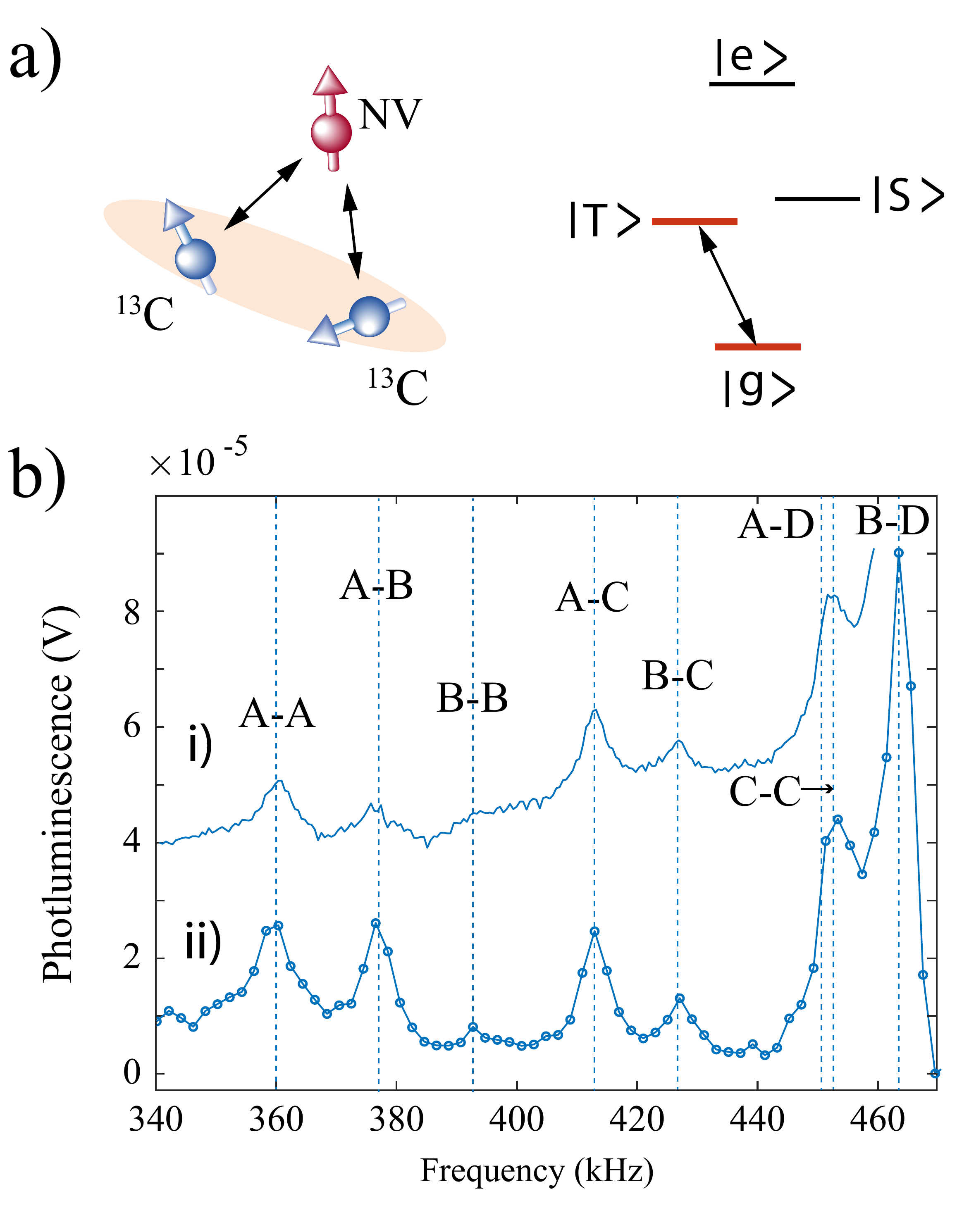}
\caption{a) Left: Schematics showing  $^{13}$C spins coupled through the NV electronic spin. Right: Energy levels of the coupled $^{13}$C spins in the $\ket{0_e}$ state. $\ket{T}$ and $\ket{S}$ denote the triplet and singlet states respectively. {The double arrow shows the transition that is detected experimentally. The transition to the singlet state is forbidden by selection rules}. b) Trace i): ODNMR spectrum in the {340-460 kHz} range, showing the coupled hetero- and homo- nuclear couplings between the $^{13}$C spins. The bottom curve ii) was obtained by lowering the RF signal by 10 dB.}
\label{pairs}
\end{figure}

{Individual nitrogen-vacancy (NV) centers can act as efficient mediators for coupling nearby nuclear spins, offering a more effective mechanism than direct nuclear–nuclear interactions, as the electronic spin of the NV provides both a stronger magnetic interaction and a controllable interface that enables coherent, long-range coupling between otherwise weakly interacting nuclear degrees of freedom}. Figure \ref{pairs}-a) illustrates how the NV center facilitates the interaction between $^{13}$C pairs. This picture can be naturally extended to larger clusters of nuclei, where the NV center mediates complex multi-spin interactions with potential applications in quantum information processing and quantum metrology \cite{Taminiau1, Bradley, Shi, Herb}.
{Such coupled spins are expected to exhibit eigenfrequencies distinct from those of uncoupled nuclei \cite{Taminiau1, Bradley} and allow access to first-order magnetic field insensitive transitions \cite{Bartling}}.

Predictions about the frequencies of the transition of coupled pairs can be performed by
diagonalizing numerically the Hamiltonian:
\begin{eqnarray}\nonumber
\frac{H}{\hbar} &=& DS_z^2 + \gamma_e \mathbf{S}\cdot\mathbf{B} + {}\gamma_n (\mathbf{I}_1+\mathbf{I}_2)\cdot\mathbf{B}\\ &+& \mathbf{S} \cdot \underline{\underline{\mathbf{A}}}_1 \cdot \mathbf{I}_1 
+\mathbf{S} \cdot \underline{\underline{\mathbf{A}}}_2 \cdot \mathbf{I}_2
\end{eqnarray}

Here, $\mathbf{I}_{1,2}$ denote the nuclear spin operators, and $ \underline{\underline{\mathbf{A}}}_{1,2}$ are the hyperfine tensors describing the interaction of nuclear spins 1 and 2 with the electronic spin $\mathbf{S}$ of the same NV center. 
{The values of $ \underline{\underline{\mathbf{A}}}_{1,2}$ are taken from Table  \ref{tab:deltaE}.} 
The frequency difference between the transitions of two coupled $^{13}$C and uncoupled $^{13}$C in the $\ket{\pm 1_e}$ states was found to be too small to be detectable are thus not listed.

In the case of two $^{13}$C nuclei that are coupled identically to the NV centers with a positive $A_{zz}$ component, the four eigenstates in the coupled basis are given by
\begin{equation}
\left\{
\begin{aligned}
|g\rangle &= \left|+\tfrac{1}{2}, +\tfrac{1}{2}\right\rangle, \\
|T\rangle &= \frac{1}{\sqrt{2}}\left(
\left|+\tfrac{1}{2}, -\tfrac{1}{2}\right\rangle
+
\left|-\tfrac{1}{2}, +\tfrac{1}{2}\right\rangle
\right), \\
|S\rangle &= \frac{1}{\sqrt{2}}\left(
\left|+\tfrac{1}{2}, -\tfrac{1}{2}\right\rangle
-
\left|-\tfrac{1}{2}, +\tfrac{1}{2}\right\rangle
\right), \\
|e\rangle &= \left|-\tfrac{1}{2}, -\tfrac{1}{2}\right\rangle.
\end{aligned}
\right.
\end{equation}

For pairs of inequivalent $^{13}$C nuclei, however, the states $\ket{T}$ and $\ket{S}$ are no longer perfectly symmetric and antisymmetric, due to their different hyperfine couplings to the NV center.

Importantly, the transition $\ket{g} \rightarrow \ket{T}$ is allowed whereas transitions to the singlet state are RF-forbidden by selection rules as shown in Fig. \ref{pairs}-a). 
Numerical simulations of the $\ket{g} \rightarrow \ket{T}$ transition frequencies for 
all possible pairs of strongly coupled $^{13}$C spins in the $\ket{0_e}$ states are listed in the table~\ref{coupledtable}.

The number of clusters of spins around each NV under the laser focal spot is calculated in the Supplementary material \cite{supp}, section IX. It is shown that there is a ratio of 1/10 between double occupancy and single occupancy of $^{13}$C nuclear spins close to NV centers in our sample. Such pairs should thus be detectable in about an hour with our setup.
Fig. \ref{pairs}-b) shows two scans taken in the region where such transitions between the $\ket{g}$ and triplet states $\ket{T}$ are expected. Most of the transitions are indeed observed, 
and reported in Tab. \ref{coupledtable}.
showing good agreement with theoretical modeling. 
{Further experiments are planned to investigate the coherence of these coupled spin pairs. At present, our sensitivity is insufficient to perform spin-echo measurements on such weak signals.}

\section{conclusion}

In conclusion, we perform high-sensitivity detection of nuclear spins in a CVD grown diamond crystal under modest magnetic fields at room temperature.   
The identification of nuclear spin environments around ensembles of NV center lays the groundwork for further dynamical control of the state of nuclear spins families, opening up the possibility to test and optimize various polarization transfer protocols with $^{13}$C spins that are currently only accessible using high-field NMR, even at to the single spin level. This includes Coherent Population Trapping (CPT) \cite{Nicolas_2018, Jamonneau, Jamonneau_2, Huillery}, pulse-pol \cite{Scheuer} and adiabatic sweeps techniques \cite{Ajoy2018, Henshaw} which are central for improving spin initialization and readout in diamond-based quantum sensors. 
Another promising avenue arises from the temperature-dependent variations observed in all spectral features \cite{Jiangfeng13C, Soshenko, Barson}, which provide a means to investigate the spin–phonon interactions of the newly detected clusters with high precision.

Importantly, our results are relevant in the context of gyroscopy. Nuclear spins have been proposed as promising candidates for the next generation of compact and precise gyroscopes \cite{Ajoy2012, ledbetter2012gyroscopes, Wang}. While the gyroscopic signal associated with $^{13}$C nuclear spin precession could in principle be enhanced using $^{13}$C-enriched samples, our findings suggest that additional experimental and theoretical efforts will be required to fully assess the feasibility and ultimate performance of such implementations, particularly in comparison with established $^{14}$N-based approaches \cite{Jarmola, Soshenko_gyro}.
\nocite{Findler, Pellet, Pagliero, Wunderlich, Shimon}

\begin{acknowledgments}
We thank Pascal Morfin for technical assistance and 
Anais Dr\'eau for useful discussions.
This project was funded within the QuantERA II Programme that has received funding from the European Union’s Horizon 2020 research and innovation programme under Grant Agreement No 101017733.
\end{acknowledgments}

\end{document}


\title{High-Sensitivity Optical Detection of Electron-Nuclear Spin Clusters in Diamond \\
 -- Supplementary materials -- }

\author{L. Chambard}
\affiliation{Laboratoire De Physique de l'\'Ecole Normale Sup\'erieure, \'Ecole Normale Sup\'erieure, PSL Research University, CNRS, Sorbonne Universit\'e, Universit\'e Paris Cit\'e , 24 rue Lhomond, 75231 Paris Cedex 05, France}

\author{A. Durand}
\affiliation{Laboratoire De Physique de l'\'Ecole Normale Sup\'erieure, \'Ecole Normale Sup\'erieure, PSL Research University, CNRS, Sorbonne Universit\'e, Universit\'e Paris Cit\'e , 24 rue Lhomond, 75231 Paris Cedex 05, France}

\author{J. Voisin}
\affiliation{Laboratoire De Physique de l'\'Ecole Normale Sup\'erieure, \'Ecole Normale Sup\'erieure, PSL Research University, CNRS, Sorbonne Universit\'e, Universit\'e Paris Cit\'e , 24 rue Lhomond, 75231 Paris Cedex 05, France}

\author{M. Perdriat}
\affiliation{Laboratoire De Physique de l'\'Ecole Normale Sup\'erieure, \'Ecole Normale Sup\'erieure, PSL Research University, CNRS, Sorbonne Universit\'e, Universit\'e Paris Cit\'e , 24 rue Lhomond, 75231 Paris Cedex 05, France}

\author{V. Jacques}
\affiliation{Laboratoire Charles Coulomb, Université de Montpellier, CNRS, Montpellier, France}

\author{G. H\'etet} 
\affiliation{Laboratoire De Physique de l'\'Ecole Normale Sup\'erieure, \'Ecole Normale Sup\'erieure, PSL Research University, CNRS, Sorbonne Universit\'e, Universit\'e Paris Cit\'e , 24 rue Lhomond, 75231 Paris Cedex 05, France}

\maketitle
\vspace{2em}
\begingroup
\renewcommand{\contentsname}{Contents} 
\tableofcontents
\endgroup
\vspace{2em}

\section{Set-up}

\begin{figure}[h]
    \centering
    \includegraphics[width=\linewidth]{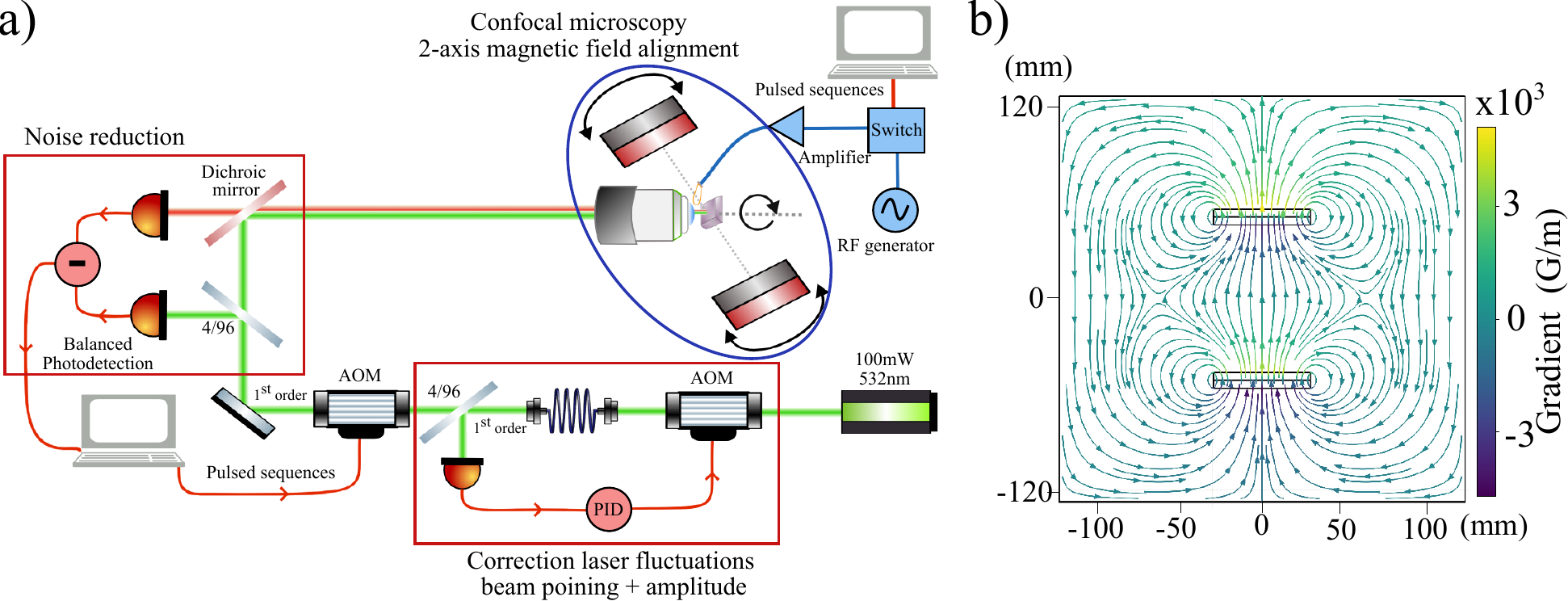}
    \caption{
    \textbf{a)} Confocal microscopy set-up. Red boxes correspond to the laser noise reduction part. The blue circle highlights the confocal microscope and the magnetic field alignment set-up with two independent rotation stages. A second AOM and a switch are used for the pulsed sequences. Two lasers are used in our measurement: Cobolt (532 nm) from Hübner and Ventus (532 nm) from Novanta. 
    \textbf{b)} Simulation (from COMSOL) of the magnetic field gradient in our set-up. Minimum gradient occurs at the center of the set-up.}
    \label{set_up}
\end{figure}

Figure~\ref{set_up}a presents the experimental setup used to perform optically detected nuclear magnetic resonance (ODNMR) measurements. The key advances that enabled the observation of weak nuclear resonances include significant noise reduction down to DC, precise control of the magnetic field direction, and a setup optimized to operate at a fixed, homogeneous magnetic field.

To reach the sensitivity required for detecting the weakest ODNMR signals, we addressed multiple noise sources, with the dominant contribution originating from laser-induced fluctuations. The laser beam is coupled into an optical fiber to mitigate both low frequency beam-pointing and amplitude instabilities. Additionally, a feedback loop controlling an acousto-optic modulator (AOM) is used to actively stabilize the green laser power.

Confocal detection is achieved using an objective lens with a numerical aperture of 0.75, corrected for chromatic aberration. This configuration enhances photoluminescence (PL) collection efficiency from the excitation volume while minimizing the effect of magnetic field gradients. The excitation power before the diamond ranges from 5~mW to 20~mW, resulting in PL signals up to 40~$\mu$W. Detection is performed using a balanced photo detector (PDB210A from Thorlabs), where the PL signal is measured relative to the green laser reference taken prior to the confocal path using a 4/96 reflective quartz plate. The power on the reference photo diode is fine-tuned using a variable optical density to optimize balancing. Under these conditions, we achieve a detection noise level of approximately 1~mV over minutes time scales, allowing us to resolve ODNMR contrasts as low as $\sim10^{-6}$ within one hour of averaging. 

\begin{figure}[h]
    \centering
    \includegraphics[width=\linewidth]{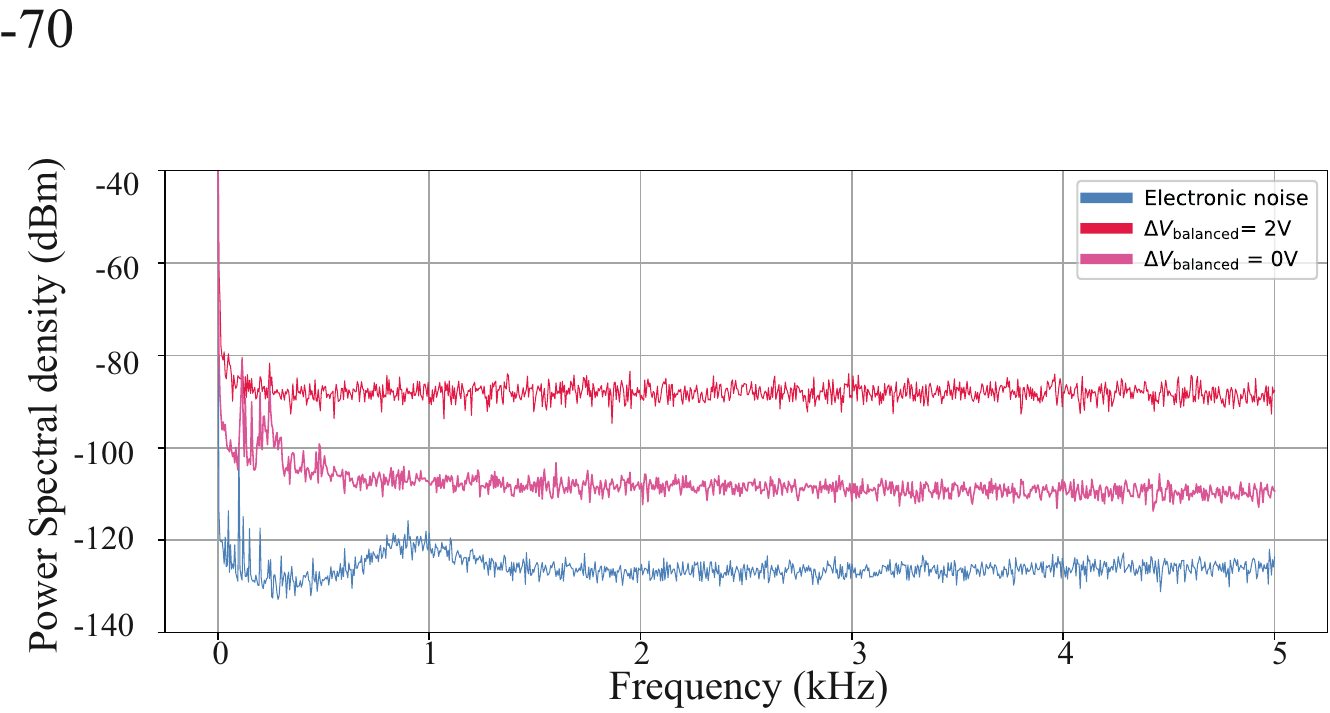}
    \caption{Power spectral density of the photoluminescence signal with and without balancing the green laser and PL signals.}
    \label{PSD}
\end{figure}

{Figure \ref{PSD} shows the power spectral density of the photoluminescence signal with ($\delta V_{\rm balanced}=0$V) and without ($\delta V_{\rm balanced}=2$V) balancing the green laser and PL signals. About 20 dBm of noise suppression is achieved from DC to 5 kHz. }

Experiments were conducted using two laser sources: a Ventus-GEM and a Cobolt Samba (DPSS), both operating at 532~nm and capable of delivering up to 200~mW of power. Most data in the main text were acquired using the Ventus-GEM laser. The only exceptions are the ODMR spectra in Figure~2, the $\ket{0_e}$ manifold data in Figures~4a and 4b, and the results in Figure~5. In the Supplementary Material, Figures~\ref{mag_calib}b and \ref{T1}b were obtained using the Cobolt Samba laser.

Magnetic field alignment is achieved using a pair of 2 inches permanent magnets to minimize field gradients. Figure~\ref{set_up}b shows a simulation of the expected magnetic field distribution in the region surrounding the diamond. The sample is mounted at the center of a goniometric system, where the field gradient is minimal—on the order of a few $0.06$ T/m along $z$ (see Fig. 1-a). With an excitation spot size along the laser beam axis estimated at $\sigma \sim100~\mu$m, this yields an inhomogeneity in the 100 ppm range corresponding to a broadening of the electronic spin transitions on the order of $\gamma_e \nabla B \sigma \approx $ 200 kHz. This is  more than a factor of three below the inverse of the inhomogeneous dephasing time $1/T_2^* \approx 650$ kHz of the electronic spins measured at low fields. 

The diamond is fixed on a high-precision rotation stage (CONEX-AG-PR100P from Newport, minimum incremental motion: 0.001$^\circ$), allowing rotation around the laser axis. A second rotation stage (CONEX-CC, minimum increment: 0.02$^\circ$) holds the magnets and enables azimuthal rotation around the diamond.

The magnetic field strength is calibrated using optically detected magnetic resonance (ODMR) in combination with numerical simulations of the electronic spin Hamiltonian, including hyperfine interaction with the $^{14}$N nuclear spin. By measuring the Zeeman splitting between the $\ket{-1_e}$ and $\ket{+1_e}$ spin sub-levels and comparing the values to simulated resonance frequencies, we extract a magnetic field magnitude of 487(1) G.

\section{Alignement of the magnetic field to the NV axis}

Efficient nuclear spin polarization via optical pumping of NV centers relies on spin mixing processes that occur at the excited-state level anti-crossing (ESLAC), with a maximum polarization near 510–520~G, when the magnetic field is aligned with the NV symmetry axis. The ESLAC condition corresponds to an avoided crossing between the $\ket{0_e}$ and $\ket{-1_e}$ sublevels of the NV excited state, which enables hyperfine-mediated mixing of electronic and nuclear spin states. This mixing is a prerequisite for transferring polarization from the NV electronic spin to nearby nuclear spins, such as $^{13}$C, $^{14}$N, or $^{15}$N.

Critically, the efficiency of this mixing—and hence of the polarization process—depends strongly on the orientation of the applied magnetic field relative to the NV axis. When the field is precisely aligned along the NV axis, the quantization axis of the spin Hamiltonian matches the symmetry axis of the defect, maximizing the overlap between spin states that participate in the ESLAC-mediated mixing. In this configuration, the secular approximation remains valid, and the perpendicular components of the hyperfine tensor efficiently drive the mixing.
For these reasons, precise alignment of the external magnetic field along the NV axis is essential for operating at the ESLAC with optimal nuclear spin polarization. In our setup, this alignment is achieved using a dual-axis goniometric mount that allows us to adjust both the diamond orientation and the magnet angle with sub-millidegree resolution (see Fig.~\ref{set_up}).
{The magnetic field is set to a value that is slightly below the optimum point to avoid cross-relaxation between NV centers and $P_1$ centers, which would otherwise reduce the NV $T_1$ time \cite{Jarmola_1}. Conversely, $P_1$ centers can in fact also assist $^{13}$C polarization \cite{pagliero, Shimon, Wunderlich}. This effect was not observed in our experiment where only the 27 $^{13}$C in close proximity to NV centers are polarized through DNP at the ESLAC (see \cite{Dreau2012} for single spin studies). For these strongly coupled spins, no polarization could be observed in the 510–520~G range in our experiment, on a par with what was observed in \cite{Jarmola_1} in the context of $^{14}$N nuclear spin polarization.}

To ensure optimal alignment of the magnetic field along the NV axis, we exploit the optical readout of $^{14}$N nuclear spin polarization via the peaks labeled $^{+1}$N$_0$ and $^{-1}$N$_0$ in the main text (see Fig.~3). These peaks arise from spin-selective optical transitions and directly reflect the nuclear polarization of the $^{14}$N nucleus associated with the NV center~\cite{Jarmola_1}. Since they can be detected within a fraction of a millisecond, they serve as an effective probe of alignment quality.
\begin{figure}
\includegraphics[width=10cm]{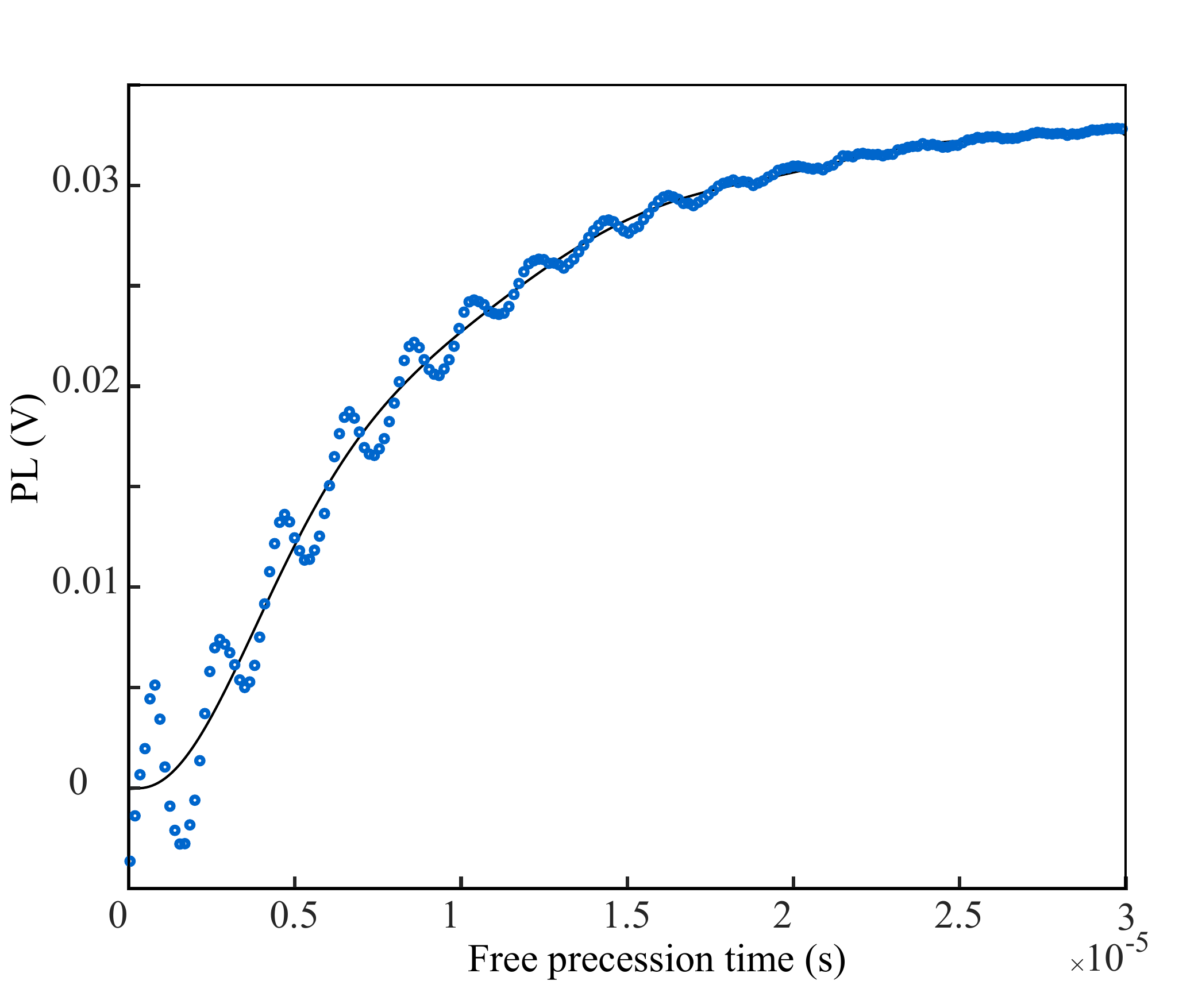}
\caption{Photoluminescence as a function of waiting time between two hard $\pi/2$ pulses and $\pi$ pulse in a spin echo sequence. The plain line is a polynomial fit to the data.}
\label{Echo}
\end{figure}
We use the relative amplitudes of these peaks to finely adjust the magnetic field direction. We define the nuclear polarization as:
\begin{equation}
    P = \frac{^{+1}\mathrm{N}_0}{^{+1}\mathrm{N}_0 + ^{-1}\mathrm{N}_0},
\end{equation}
where $^{+1}\mathrm{N}_0$ and $^{-1}\mathrm{N}_0$ denote the amplitudes of the corresponding peaks. 
 A two-dimensional map of the nuclear polarization as a function of rotation stage angles is shown in Fig.~\ref{mag_calib}.a).
Using this method, we achieve an angular alignment precision better than $0.01^\circ$. This high degree of control is essential to maintain efficient polarization transfer at the ESLAC and to ensure reproducible measurements across experiments.

\section{ESEEM}

Electron Spin Echo Envelope Modulation (ESEEM) is a pulsed electron spin resonance technique used to probe weak hyperfine and quadrupolar interactions between an electron spin and nearby nuclear spins. It is based on the observation that in the presence of weakly coupled nuclei, the amplitude of the Hahn echo signal exhibits modulations as a function of the interpulse delay.

These modulations arise because the nuclear spin precession frequency depends on the electron spin state due to the hyperfine interaction. During the two free evolution periods of the Hahn echo sequence ($\pi/2$ -- $\tau$ -- $\pi$ -- $\tau$ -- $\pi/2$), the nuclear spin accumulates a phase which partially cancels or reinforces depending on the electron spin path, leading to an echo amplitude modulated by oscillations at characteristic frequencies. This is called an Electron Spin Echo Envelope Modulation (ESEEM). For a nuclear spin 1/2 such as that of the $^{13}$C under a magnetic field aligned with the NV center axis, the modulation frequency equals the Larmor frequency $|\gamma_n B|$.

Fig. \ref{Echo} shows a plain spin echo signal obtain by driving the electronic spin transition at 1.506 GHz. The PL is plotted as a function of the precession time $\tau$.

In the main text, the oscillating signal is subtracted to let the $^{13}$C Larmor precession appear clearly and to facilitate Fourier analysis. While ESEEM spectra can provide valuable information about the hyperfine tensor under carefully chosen magnetic field angles, the resolution $\Delta \nu$ of the ESEEM spectrum is limited by the coherence time $T_2$ (via $\Delta \nu \approx 1/(2T_2$)). 

\section{Origin of the $^{13}$C transitions in the  $\pm 1_e$ states}

\begin{figure}
\includegraphics[width=15cm]{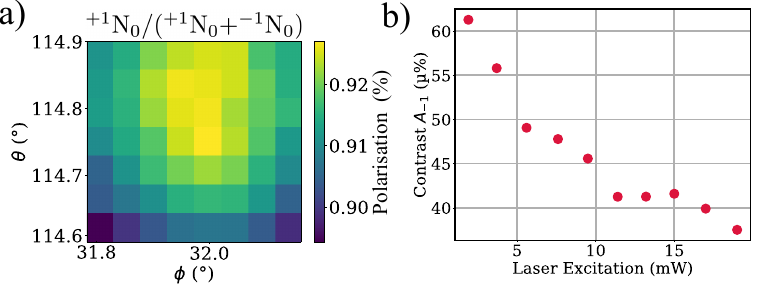}
\caption{Polarization a) Evolution of the polarization defined by the ratio of the resonance amplitude in the manifold $\ket{0_e}$ with the rotation stages angles. b) Evolution of the contrast of the resonance A$_{-1}$ as a function of the green excitation power measured just before the confocal microscope.}
\label{mag_calib}
\end{figure}

The peaks associated with the hyperfine interaction with $^{13}$C in the $\ket{\pm 1_e}$ manifold (Fig. 5 in the main text) are observed without applying microwave excitation to the NV magnetic states. These resonances appear while the laser continuously pumps the NV center into the $\ket{0_e}$ state.

To better understand the observation of resonances linked to the hyperfine interaction with $^{13}$C in the $\ket{\pm 1_e}$ manifold, we analyze in Fig.~\ref{mag_calib}.b) the evolution of the contrast of the resonance $A_{-1}$ as a function of the green excitation power. We observe that the contrast decreases with increasing laser power, which indicates that it is directly related to the residual population remaining in the $\ket{-1_e}$ state due to the ineffective optical pumping of the NV center's spin to the $\ket{0_e}$ state.

The optimum laser power to observe these features must be found by compromising sensitivity (which scales as the square root of the power for a shot noise limited detection scheme) to contrast.  

\section{Electronic spin relaxation time $T_1$}

In Figure 8.a) of the main text, using Ramsey oscillations on the nuclear spin, we extracted a $T_2^*$ value of 864.2$\mu$s for the $^{14}$N. Since our readout of the nuclear spin state is performed via coupling to the electronic spin state, we need to determine whether the electronic relaxation time (the electron $T_1$ time) limits the nuclear $T_2^*$.

\begin{figure}
\includegraphics[width=16cm]{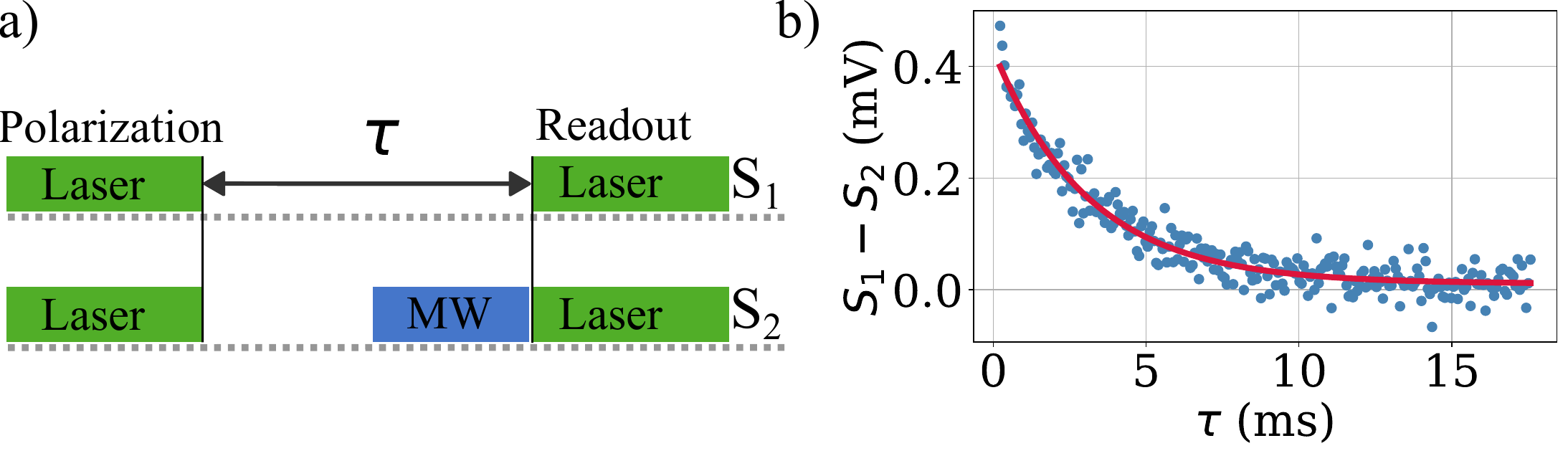}
\caption{a) Pulse sequence used for  $T_1$ determination. The microwave frequency corresponds to the electronic transition $\ket{0_e}\rightarrow\ket{-1_e}$ for the NV centers aligned with the magnetic field.
b) Evolution of the difference between signals $S_1$ and $S_2$ as a function of the delay time $\tau$. The blue points represent the experimental data, and the red line is an exponential decay fit.}
\label{T1}
\end{figure}

Figure \ref{T1} shows the pulse sequence used and the data obtained to determine the electronic spin $T_1$. We take the difference between signals $S_1$ and $S_2$ in order to eliminate other effects that occur when the laser is turned on, such as photo-ionization of NV$^-$ to NV$^0$. The microwave pulse only affects the electronic spin state, so the signal difference provides direct information about the relaxation time. With an exponential decay fit we extract an electronic relaxation time $T_1$=3.1(1)ms \cite{Pellet}. This value is longer than the $T_2^*$ of the $^{14}$N nuclear spins, indicating that our Ramsey measurements are not limited by the electronic relaxation time.

The $T_2^*$ value is thus likely related to fluctuating magnetic noise originating from P$_1$ or NVH$^-$ centers, which are present at the ppm level in our sample \cite{Findler}. 

\section{Hamiltonian of the $^{13}$C-NV coupling}
The general Hamiltonian for the electronic spin ${\bf S}$ of an NV center coupled to the nuclear spin ${\bf I}$ of a $^{13}$C in a magnetic field ${\bf B}$ reads:
$$\frac{H}{\hbar} = DS_z^2 + \gamma_e \mathbf{B}\cdot \mathbf{S} + \gamma_n\mathbf{B}\cdot \mathbf{I} + \mathbf{S} \cdot \mathbf{A} \cdot \mathbf{I}$$

where $\gamma_e=2.8025~\rm MHz/G$ and $\gamma_n=-1.07~\rm kHz/G$ are the gyromagnetic ratios 
of the electronic and of the $^{13}$C nuclear spins respectively.
Writing the spin operator components in terms of creation and annihilation operators, we obtain 

We obtain
$$
\begin{aligned}
    H/\hbar &= DS_z^2 + \gamma_e \mathbf{B}\cdot \mathbf{S} + \gamma_n\mathbf{B}\cdot \mathbf{I} \\
  &+ A_{zz} S_z I_z + \frac{1}{2}  S_z I_+ (A_{zx} - i A_{zy}) + \frac{1}{2}  S_z I_- (A_{zx} + i A_{zy}) \\
  &+ \frac{1}{2} S_+ I_z (A_{xz} - i A_{yz}) + \frac{1}{2} S_- I_z (A_{xz} + i A_{yz}) \\
  &+ \frac{1}{4} S_+ I_+ (A_{xx} - A_{yy} - 2 i A_{xy} ) \\
  &+ \frac{1}{4} (S_+ I_- +S_- I_+) (A_{xx} + A_{yy} ) \\
  &+ \frac{1}{4} S_- I_- (A_{xx} - A_{yy} + 2 i A_{xy}).
\end{aligned}
$$
The dipole-dipole interaction between a carbon 13 nucleus and the NV electrons can be expressed in a matrix form as
$$
\hat H_{dd}=\mathbf{S} \cdot \mathbf{A}_d \cdot \mathbf{I}
$$
where $\mathbf{A}_d$ is the symmetric zero-trace matrix with elements:
$$
[\mathbf{A}_d]_{ij}=\xi(\hat r)(\delta_{ij}-3 \hat{n}_i \hat{n}_j)
$$
and 
$\xi(\hat{r})\equiv -A_d=\frac{\mu_0\gamma_e \gamma_n}{4\pi \hat{r}^3}$.\\

$n_i$ are the coordinates of the unit vector $ \bf{\hat n}$ that joins the spins, with origin on the vacancy (where the electronic density is the largest in the NV ground state \cite{DOHERTY20131}).
We write $\theta$ and $\phi$, the polar and azimuthal angles that determine the coordinates of $ \bf{\hat n}$.
Adding an identical contact term $A_c$ to the diagonal elements \cite{Duarte}, we get to

\begin{align}
    A_{xx} &= A_c - A_d (1 - 3 \sin^2\theta \cos^2\phi), \\
    A_{yy} &= A_c - A_d (1 - 3 \sin^2\theta \sin^2\phi), \\
    A_{zz} &= A_c - A_d (1 - 3 \cos^2\theta), \\
    A_{xy} &= A_{yx} = 3 A_d \sin^2\theta \cos\phi \sin\phi, \\
    A_{xz} &= A_{zx} = 3 A_d \sin\theta \cos\theta \cos\phi, \\
    A_{yz} &= A_{zy} = 3 A_d \sin\theta \cos\theta \sin\phi.
\end{align}

We have:
\begin{align}
    A_{xx} - A_{yy} \pm 2i A_{xy} &= 3 A_d \sin^2\theta e^{\pm i 2\phi}, \\
    A_{xz} \pm i A_{zy} &= 3 A_d \cos\theta \sin\theta e^{\pm i \phi}.
\end{align}

The hyperfine interaction can then be written as
\begin{align}
  \mathbf{S} \cdot \mathbf{A} \cdot \mathbf{I} &= S_z I_z A_{\parallel}  + \frac{1}{4}(2A_c - A_d(2 - 3 \sin^2\theta))(S^+ I^- + S^- I^+) \nonumber \\
    &\quad + \frac{3}{4} A_d \sin^2\theta (S^+ I^+ e^{-2i\phi} + S^- I^- e^{+2i\phi}) \nonumber \\
    &\quad + \frac{3}{2} A_d \cos\theta \sin\theta ((S^+ I_z + S_z I^+) e^{-i\phi} + (S^- I_z + S_z I^-) e^{+i\phi}).
\end{align}

We now define
\begin{align}
A_{\parallel}&=A_{zz},\\
    A_{\perp} &= \frac{2A_c + A_d (1 - 3 \cos^2\theta)}{2}=\frac{3 A_c-A_{\parallel}}{2}=\frac{A_{xx}+A_{yy}}{2}, \\
    A_{\rm  ani} &= 3 A_d \cos\theta \sin\theta, \\
    A'_{\perp} &= \frac{3 A_d \sin^2\theta}{2}=\frac{A_{xx}-A_{yy}}{2},
\end{align}
so that 
\begin{align}
    \mathbf{S} \cdot \mathbf{A} \cdot \mathbf{I} &= S_z I_z A_{\parallel} \nonumber \\
    &\quad + \frac{A_{\perp}}{2} (S^+ I^- + S^- I^+) \nonumber \\
    &\quad + \frac{A'_{\perp}}{2} (S^+ I^+ e^{-2i\phi} + S^- I^- e^{+2i\phi}) \nonumber \\
    &\quad + \frac{A_{\text{ani}}}{2} ((S^+ I_z + S_z I^+) e^{-i\phi} + (S^- I_z + S_z I^-) e^{+i\phi}).
\end{align}
In the following, we use the fact that $A_\perp'\ll A_\perp$ because $A_{xx}\approx A_{yy}$ \cite{PhysRevB.92.115206}. We then obtain the $^{13}$C Hamiltonian shown in the main text. We will come back to this approximation in the next section.

The sign of $A_c$ depends on the $^{13}$C family around the NV center. 
For the strongly coupled families A, B, C and D that we study in this paper, the dipolar interaction terms $\propto A_d$ are smaller than 
$A_c$. Therefore the signs of $A_\perp$ and $A_\parallel$ are the same.

Note also that $A_d >0$. 
For the families A,B, C, and D, $\theta \in [0,\pi/2]$ with our choice of coordinate system (see Fig. 1 in the main text), therefore $A_{\rm ani}>0$ for these four families. 

\section{Second-order perturbation theory of degenerate states}\label{SOT}


To explain the energy level repulsion observed between the states $\ket{0_e, +1/2}$ and $\ket{0_e, -1/2}$, we employ second-order perturbation theory perturbation theory of degenerate states. This approach allows us to describe second-order couplings mediated by virtual transitions through intermediate states outside the $\ket{0_e}$ manifold. This approach allows to derive an effective Hamiltonian projected onto the $\ket{0_e}$ subspace, accounting for the observed shifts. We write the total Hamiltonian as $H = H_0 + V$, where:
$$
 \begin{aligned}
     H_0 &= DS_z^2 +\gamma_eS_zB_z+\gamma_n I_zB_z + S_zI_zA_{\parallel},\\
     V &= \frac{A_\perp}{2}(S^+I^- + S^-I^+) +\frac{A_{\rm  ani}}{2}((S^+I_z+S_zI^+)e^{-i\phi} + (S^-I_z+S_zI^-)e^{+i\phi})+\gamma_e B_xS_x + \gamma_nB_x I_x.
 \end{aligned}
$$
We set $\hbar$ to 1 here to simplify notations.
Because of the small typical values of the hyperfine tensor and of $B_x$ compared to the splittings between eigen-states of $H_0$ coupled by $V$, we can consider $V$ as a perturbation. Second order perturbation theory states that the matrix elements of the effective Hamiltonian in the sub-space $\ket{0_e}$ read: 
 $$
 \bra{i}H_{eff}^{\ket{0_e}}\ket{j}= \bra{0_e}H\ket{0_e}\otimes Id(2) + \bra{0_e,i}V\ket{0_e,j} + \frac{1}{2}\sum_{k,\gamma\in\{+1_e, -1_e\}}\bra{0_e,i}V\ket{\gamma,k}\bra{\gamma,k}V\ket{0_e,j}\Big(\frac{1}{E_{0,i}-E_{\gamma,k}}+\frac{1}{E_{0,j}-E_{\gamma,k}}\Big)
 $$
 
This expression includes contributions from direct level repulsion between nearly degenerate states within the same subspace $\ket{0_e}$, as well as indirect energy shifts mediated by virtual transitions through higher-energy intermediate states. These are depicted by arrows in the level scheme shown in Fig.~\ref{couplings}-a).\\

The terms arising from first order perturbation are: 
$$ 
\begin{aligned}
    \bra{0_e,-1/2}V\ket{0_e,+1/2} &= \frac{\gamma_n B_x}{2}\\
    \bra{0_e,+1/2}V\ket{0_e,-1/2} &= \frac{\gamma_n B_x}{2}
\end{aligned}
$$


Neglecting $A_\parallel$ and $\gamma_n B_z$ compared to $D\pm \gamma_e B_z$, the diagonal terms read: 
$$
\begin{aligned}
    |\bra{0_e, 1/2}V\ket{1_e, +1/2}|^2\Big(\frac{2}{E_{0,1/2}-E_{1,+1/2}}\Big) &=  -\frac{1}{D+\gamma_eB_z}\times|\frac{A_{\rm ani}e^{i\phi}}{2}+\gamma_eB_x|^2\\
    |\bra{0_e, 1/2}V\ket{-1_e, +1/2}|^2\Big(\frac{2}{E_{0,1/2}-E_{-1,+1/2}}\Big) &=  -\frac{1}{D-\gamma_eB_z}\times|\frac{A_{\rm ani}e^{-i\phi}}{2}+\gamma_eB_x|^2\\
    |\bra{0_e, 1/2}V\ket{1_e, -1/2}|^2\Big(\frac{2}{E_{0,1/2}-E_{1,-1/2}}\Big) &=  -\frac{A_\perp^2}{D+\gamma_eB_z}\\
    |\bra{0_e, -1/2}V\ket{-1_e, +1/2}|^2\Big(\frac{2}{E_{0,-1/2}-E_{-1,+1/2}}\Big) &=  -\frac{A_\perp^2}{D-\gamma_eB_z}\\ 
    |\bra{0_e, -1/2}V\ket{1_e, -1/2}|^2\Big(\frac{2}{E_{0,-1/2}-E_{1,-1/2}}\Big) &=  -\frac{1}{D+\gamma_eB_z}\times|\frac{-A_{\rm ani}e^{i\phi}}{2}+\gamma_eB_x|^2\\
    |\bra{0_e, -1/2}V\ket{-1_e,-1/2}|^2\Big(\frac{2}{E_{0,-1/2}-E_{-1,-1/2}}\Big) &=  -\frac{1}{D-\gamma_eB_z}\times|\frac{-A_{\rm ani}e^{-i\phi}}{2}+\gamma_eB_x|^2\\
\end{aligned}
$$

The off-diagonal terms are: 
$$
\begin{aligned}
    \bra{0_e, -1/2}V\ket{-1_e, +1/2}\bra{-1_e, +1/2}V\ket{0_e, +1/2}\Big(\frac{1}{E_{0,-1/2}-E_{-1,1/2}}+\frac{1}{E_{0,1/2}-E_{-1,1/2}}\Big) &=  \frac{-A_\perp}{D-\gamma_eB_z}(\frac{A_{\rm ani}e^{i\phi}}{2}+\gamma_eB_x)\\
    \bra{0_e, -1/2}V\ket{1_e, -1/2}\bra{1_e, -1/2}V\ket{0_e, +1/2}\Big(\frac{1}{E_{0,-1/2}-E_{1,-1/2}}+\frac{1}{E_{0,1/2}-E_{1,-1/2}}\Big) &=  \frac{-A_\perp}{D+\gamma_eB_z}(\frac{-A_{\rm ani}e^{i\phi}}{2}+\gamma_eB_x)\\.
\end{aligned}
$$

From this we can give the expression for each component of the effective Hamiltonian. Writing $h_{ij}= \bra{i}H_{eff}^{\ket{0_e}}\ket{j}$, we find 

$$
\begin{aligned}
    h_{11} &= \frac{\gamma_nB_z}{2} + \frac{A_\perp^2\gamma_eB_z}{2\Delta^2} -\frac{DA_{\rm ani}\cos\phi\gamma_eB_x}{\Delta^2} - \frac{D}{2\Delta^2}(2\gamma_e^2B_x^2+\frac{A_{\rm ani}^2}{2}+A_\perp^2)\\
\end{aligned}
$$
and
$$
\begin{aligned}
    h_{22} &= -\frac{\gamma_nB_z}{2} -\frac{A_\perp^2\gamma_eB_z}{2\Delta^2} +\frac{DA_{\rm ani}\cos\phi\gamma_eB_x}{\Delta^2} - \frac{D}{2\Delta^2}(2\gamma_e^2B_x^2+\frac{A_{\rm ani}^2}{2}+A_\perp^2)
\end{aligned}
$$
where we have introduced the quantity $\Delta^2= D^2-\gamma_e^2B_z^2$.
It can be rewritten  as: 
$$
\begin{aligned}
    h_{11} &= \frac{1}{2}(\gamma_nB_z +\nu +C)\\
    h_{22} &= \frac{1}{2}(-\gamma_nB_z -\nu +C)
\end{aligned}
$$
where
$$
\begin{aligned}
    \nu &= - \frac{2DA_{\rm ani}\cos\phi\gamma_eB_x}{\Delta^2} +\frac{\gamma_eB_zA_\perp^2}{\Delta^2}\\
    C &= - \frac{D}{\Delta^2}(2\gamma_e^2B_x^2+\frac{A_{\rm ani}^2}{2}+A_\perp^2)
\end{aligned}
$$
We find here the expression of $\nu$ which is in the main text. The off-diagonal energy terms can be shifted by $-C$ without loss of generality to get the same effective Hamiltonian $H_{\rm eff}^{\ket{0_e}}$ as the one in the main text. \\

For the off-diagonal terms, we have contributions from both first- and second-order perturbations. We denote the latter as $h_\perp$ in the main text. We have: 

$$
\begin{aligned}
     h_{12} &=  \frac{\gamma_n B_x}{2}  - \frac{A_\perp}{2\Delta^2}(2D\gamma_eB_x + A_{\rm ani}e^{i\phi}\gamma_eB_z)
\end{aligned}
$$
From this we can deduce the expression of $h_\perp$:
$$
\begin{aligned}
    h_\perp&=    - \frac{A_\perp}{\Delta^2}(2D\gamma_eB_x + A_{\rm ani}e^{i\phi}\gamma_eB_z)
\end{aligned}
$$

Let us come back to the $A_\perp' \ll A_\perp$ approximation made earlier.
A straightforward calculation shows that including the term proportional to $A_\perp'$ in the Hamiltonian, implies making the replacement $A_\perp^2 \rightarrow A_\perp^2 -A_\perp'^2 $ in $\nu$ and $A_\perp \rightarrow A_\perp -A_\perp' $ in $h_\perp$. The contact term being dominant for all the strongly coupled spins of this study, $A_\perp' \ll A_\perp$ so since $A_\perp'$ appears only next to $A_\perp$, neglecting $A_\perp'$ is justified for these families. {The good agreement with theoretical calculations performed in \cite{Ivady} (see Table III in main text)  supports the validity of these approximations.} 

\section{$^{14}$N transitions in the NV$^-$ and NV$^0$ states}

Because of the large quadrupole moment of ${^{14}}$N, the NMR transition frequencies for the ${^{14}}$N nuclear spin in the NV$^-$ and NV$^0$ states can be found using second order perturbation theory of non-degenerate states. We obtain the following expressions for the NMR transition frequencies in the NV$^-$ state, to lowest order in $A_\perp/(D\pm \gamma_e B)$:

\begin{align*}
f(^{+1}N_0) &\approx |Q| - \gamma_n^{(N)} B - \frac{A_\perp^2}{D - \gamma_e B} \\
f(^{-1}N_0 )&\approx |Q| + \gamma_n^{(N)} B - \frac{A_\perp^2}{D + \gamma_e B} \\
f(^{+1}N_{-1}) &\approx |Q| - |A_\parallel| - \gamma_n^{(N)} B \\
f(^{-1}N_{-1} )&\approx |Q| + |A_\parallel| + \gamma_n^{(N)} B + \frac{A_\perp^2}{D - \gamma_e B} \\
f(^{+1}N_{+1})&\approx |Q| + |A_\parallel| - \gamma_n^{(N)} B + \frac{A_\perp^2}{D + \gamma_e B} \\
f(^{-1}N_{+1})&\approx |Q| - |A_\parallel| + \gamma_n^{(N)} B
\end{align*}
 We obtain the same results as in \cite{lourette2023}, if one takes into account our different sign convention for the gyromagnetic factors.

A similar procedure yields 
\begin{align*}
f(^{+1}N_{-1/2}) &\approx |Q_0| + \frac{A^0_\parallel}{2} - \gamma_n^{(N)} B \\
f(^{-1}N_{-1/2} )&\approx |Q_0| - \frac{A^0_\parallel}{2} + \gamma_n^{(N)} B  - \frac{(A^0_\perp)^2}{2\gamma_e B} \\
f(^{+1}N_{+1/2})&\approx |Q_0| - \frac{A^0_\parallel}{2} - \gamma_n^{(N)} B + \frac{(A^0_\perp)^2}{2\gamma_e B} \\
f(^{-1}N_{+1/2})&\approx |Q_0| + \frac{A^0_\parallel}{2} + \gamma_n^{(N)} B 
\end{align*}
for the transition frequencies in the $S=1/2$ ground state of the neutral NV, to lowest order in $A^0_\perp/\gamma_e B$. 
These expressions are in agreement with numerical simulations. The transition frequencies we measure are consistent with the reported values of the hyperfine coupling in a $S=1/2$ state and not with for an $S=3/2$ state where nine features should have been detected at higher frequencies \cite{Felton_neutral}. 

One can get an order of magnitude estimate of the hyperfine constants, including their signs, in the neutral state, using the ratio between the gyromagnetic ratios for both isotopes. These are $\gamma_n^{(N)}=-0.3077$ MHz/T and $\gamma_n^{(N)}=0.4316$ MHz/T for the isotope 14 and 15 respectively. 
The norm of $|^{14} A_\parallel|\approx |6.06|$ MHz for the neutral NV, while $|^{15} A_\parallel|\approx |8.484| $ MHz \cite{Waldherr}. 
The ratio of their norm equal the ratio of the norm of the gyromagnetic ratios $\approx 1.40$ and their sign is different.\\


 Hyperfine components have been reported to be negative in the $^{15}$N isotope in Felton {\it et al} \cite{Felton_neutral}. The components of the $^{14}$N diagonal components of the hyperfine tensor are then positive in the NV$^0$ state. 
Note that the perpendicular components in the $S=1/2$ state of the NV$^0$ state has not been reported so far.


\begin{figure}
\includegraphics[width=16cm]{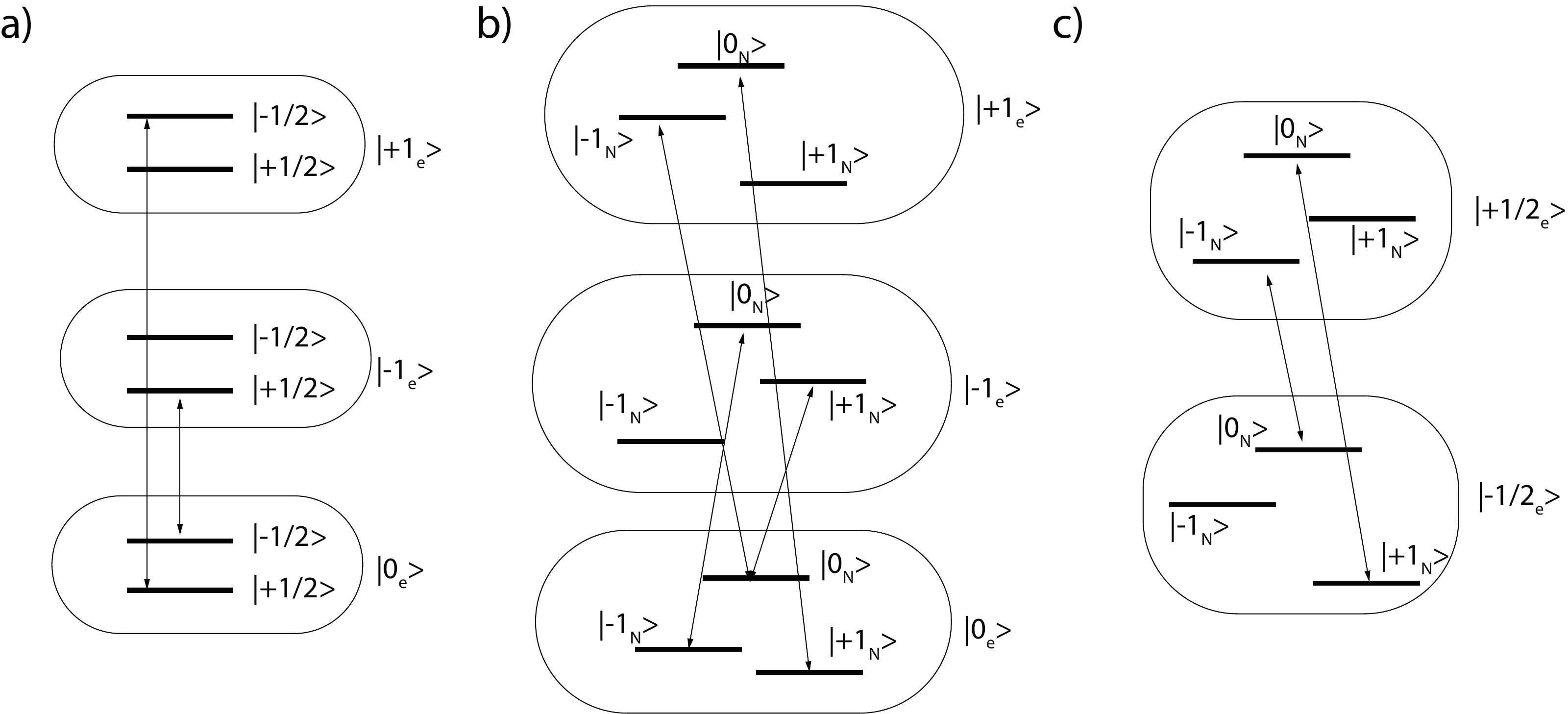}
\caption{Level schemes showing the various second order couplings in the hyperfine coupling in three different scenarios: 
a) Nuclear spin $I=1/2$ and electronic spin $S=1/2$.
b) Nuclear spin $I=1$ and electronic spin $S=1$.
c) Nuclear spin $I=1$ and electronic spin $S=1/2$.
Electronic spins states are indicated on the right of each circled nuclear spin manifold. 
}
\label{couplings}
\end{figure} 
The transverse second-order coupling of the nuclear spins to the electronic spins is mediated by the flip-flop term $S^+ I^- + h.c.$, which leads to state repulsion between the coupled levels represented in the Fig. \ref{couplings}.
The second order coupling to the electronic states thus tends to lower/increase the nuclear transition frequency in the ground/excited electronic state as can be checked in the equations above.\\

In order to extract experimentally the various hyperfine and quadrupolar tensor components shown in table I of the main text, we make use of the following combinations of these four equations.

\begin{align*}
|Q_0|&=\frac{1}{4}\Big[f(^{+1}N_{-1/2})+f(^{-1}N_{-1/2} )+f(^{+1}N_{+1/2})+f(^{-1}N_{+1/2})\Big]\\ 
A^0_\parallel&=\frac{1}{2}\Big[f(^{-1}N_{-1/2})-f(^{1}N_{-1/2} )+f(^{+1}N_{+1/2})-f(^{-1}N_{+1/2})\Big]\\
A^0_\perp&=\sqrt{\gamma_e B \Big[f(^{1}N_{-1/2})+f(^{-1}N_{-1/2} )-f(^{+1}N_{+1/2})-f(^{-1}N_{+1/2})\Big]}.
\end{align*}

Note that, quantifying a second order coupling term, $A_\perp$ is obtained by subtracting two 
very close values. The uncertainty is thus much larger than that of the $Q$ and $A_\parallel$ values.

\section{$^{13}$C NMR contrast}

In this section we provide theoretical estimates of the expected NMR contrast and compare them to the experimental observations. 

\subsection{Number of single and double $^{13}$C occupancies}

Here, we estimate the probabilities of having a single and two $^{13}$C nuclei around an NV center.

To this end, let us consider a lattice of sites in which only some are occupied by a carbon-13 atom. Each site in the lattice is occupied independently with probability \( p = 0.0107 \), {\it i.e.}, with a random Bernoulli-type occupation reflection the isotopic concentration of carbon 13 atoms.
A volume element \( V \) is placed randomly on the lattice. We assume that \( V \) contains exactly \( N = 6+3+3+6=18 \) lattice sites corresponding to the strongly coupled carbon 13 spins in this study.

Let \( X \) be the random variable describing the number of occupied sites within volume \( V \). This number follows a binomial distribution with parameters N and p with probability law:
\[
\mathbb{P}(X = k) = \binom{N}{k} p^k (1 - p)^{N - k}
\]
Let \( k = 1 \). Then:
\[
\mathbb{P}(X = 1) = \binom{18}{1} \cdot 0.01 \cdot (1 - 0.01)^{17} = 0.15  
\]
\[
\mathbb{P}(X = 2) = \binom{18}{2} \cdot (0.01)^2 \cdot (1 - 0.01)^{16} = 0.013
\]
\[
\mathbb{P}(X = 3) = \binom{18}{3} \cdot (0.01)^3 \cdot (1 - 0.01)^{15} = 7\times 10^{-4}
\]

From this calculation, we find that the probability of an NV being surrounded by a single $^{13}\mathrm{C}$ atom from the A, B, C, D family is larger by a factor of $0.15/0.013 \approx 11.5$ compared to the probability of being surrounded by two $^{13}\mathrm{C}$ spins from these families.

\subsection{NMR contrast}

Here, we provide an expression of the expected ratio between the NMR contrasts of the transitions corresponding to the $^{13}$C families in the $\ket{0_e}$ state of the NV$^-$ center.


Let us first rewrite $H_{eff}^{\ket{0_e}}$ found in section 
\ref{SOT} as 
$$
H_{eff}^{\ket{0_e}}=  (\omega_0+\Omega')  \sigma_z +  ( {\Omega}  +  \omega_\perp ) \sigma_x,
$$
where $\sigma_{x,z}$ are Pauli matrices and

\bea
\omega_0&=&\frac{1}{2} \gamma_n^\parallel B_z \\
\Omega'&=& - \frac{2DA_{\rm ani}\cos\phi}{2\Delta^2} \gamma_e B_x\\
\Omega&=& \frac{1}{2} \gamma_n^\perp B_x\\
\omega_\perp&=& - \frac{A_\perp}{2\Delta^2}A_{\rm ani}e^{i\phi}\gamma_eB_z,
\eea

where 

\bea
\gamma_n^\parallel&=& \gamma_n(1-\frac{\gamma_e}{\gamma_n} \frac{A_\perp^2}{\Delta^2})\\
\gamma_n^\perp&=&\gamma_n(1-\frac{\gamma_e}{\gamma_n} \frac{2 D }{\Delta^2}A_\perp ),
\eea

Taking an RF field aligned with the $x$ axis and neglecting static transverse magnetic fields, the Hamiltonian can thus be written as  
\bea
\tilde{H}_{eff}^{\ket{0_e}}(t) = ( \omega_0+\Omega' \cos(\omega t))  \sigma_z +  ( {\Omega}\cos(\omega t)  +  \omega_\perp ) \sigma_x,
\eea

Moving into a rotating frame at the RF drive frequency, and using a rotating wave approximation, it becomes,
$$
H'= \frac{\Delta}{2}\sigma_z + \frac{\Omega}{2}  \sigma_x 
$$
where 
$\Delta=\omega-\omega_0$.
The hyperfine enhanced Rabi frequency  
$\Omega=\gamma_n^\perp B_x$ then has the same form as that of the $^{14}$N coupled to NV center \cite{chen_prb}. 

Writing the amplitude $S_A$ of the NMR signal from the family $A$, we define the ratio
$$
R^i=\frac{S^i}{S^A}, \quad {\rm where} \quad S^i \propto p^i { |\Omega^i|}^2N^i_{\rm sites}.
$$
Here $p^i$ is the nuclear spin polarizations for families $i=$A,B,C and D -- which can be obtained from \cite{Dreau2012} --, ${ \Omega^i}$ are the family-dependent Rabi frequencies and $N_{\rm sites}^i$ is the number of sites in each family. This calculation assumes that the spins are not saturated by the microwave drive, which is the case of the low power spectrum shown in the main text (Fig. 6b). 

The three theoretical values ($R^i_{\mathrm{th}}$) and the experimentally measured values ($R^i_{\mathrm{exp}}$) of this ratio  shows good overall agreement. This adds weight to the hypothesis that the NMR peaks in the $\ket{0_e}$ state correspond to the A,B, C and D families, in ascending order of frequency. Quantitative comparison, including the differing ODNMR read-out efficiency for each family, will be the subject of future work. 



%

%